\documentclass[a4paper,11pt]{article}
\usepackage[utf8]{inputenc}
\usepackage{jcappub}

\pdfoutput=1
\usepackage{subfig}
\usepackage{epsfig}
\usepackage{graphicx}
\usepackage{color}
\usepackage{amssymb}
\usepackage{amsmath}
\allowdisplaybreaks
\usepackage{mathtools}
\usepackage{url}
\usepackage{natbib}
\usepackage[normalem]{ulem}
\usepackage{afterpage}
\usepackage{float}
\usepackage{ulem}
\usepackage{verbatim}

\newcommand{\id}{\textrm{1\kern-.25em I}}

\def \HH {\mathcal{H}}
\def \ndv {v_{\parallel}}

\def \bn {\boldsymbol{n}}

\def \bv {\boldsymbol{v}}

\def \eH {\epsilon_\mathcal{H}}

\title{Gravitational Redshift from Galaxy Clusters -- a Relativistic Approach}
\author[a]{Enea Di Dio,}
\author[a]{Sveva Castello}
\author[a]{and Camille Bonvin}

\date{\today}

\affiliation[a]{Universit\'e de Gen\`eve, D\'epartement de Physique Th\'eorique and Centre for Astroparticle Physics,
24 quai Ernest-Ansermet, CH-1211 Gen\`eve 4, Switzerland}

\emailAdd{enea.didio@unige.ch}
\emailAdd{sveva.castello@unige.ch}
\emailAdd{camille.bonvin@unige.ch}

\abstract{

The light that we receive from clusters of galaxies is redshifted by the presence of the clusters' gravitational potential. This effect, known as gravitational redshift, was first detected from a sample of stacked clusters in 2011, by taking redshift differences between the centre of each cluster and the respective member galaxies.  However, the interpretation of this result was later challenged by several studies, which emphasised the possible influence of additional kinematic effects on the observed signal, like the transverse Doppler effect. In this work, we present the first derivation of all such effects within a relativistic framework, accurate to third order in the weak-field approximation. This framework allows us to correctly capture the hierarchy of terms on the scale of clusters and at the same time account for all relativistic effects. We compare our result with previous literature and show that some terms of the same order of the transverse Doppler effect were not properly included, leading to an overestimation of the kinematic contamination. In particular, we do not find any contribution arising from the so-called light-cone effect and obtain a larger correction due the motion of the central galaxy.
Our derivation is independent of the Euler equation, providing a straightforward framework to test the weak equivalence principle.
}

\begin{document}

\maketitle

\section{Introduction}

Gravitational redshift is an effect first predicted by Einstein~\cite{Einstein1907}, describing the frequency shift experienced by photons when they escape a gravitational potential. 
This shift is due to the fact that time does not pass at the same rate inside and outside a gravitational potential. Various experiments have measured this effect, from the laboratory up to cosmological scales, see e.g.~\cite{Pound:1959PhR, Lopresto:1991oxy, Alam:2017izi}. Being sensitive to the depth of the temporal gravitational potential, gravitational redshift provides a key test of the fundamental interaction between gravity and matter. Hence, it is natural to turn to the largest gravitationally bound objects in the Universe, i.e.~galaxy clusters, to look for a detection.

\begin{figure}[th!]
 \centering
	\includegraphics[width=0.7\columnwidth]{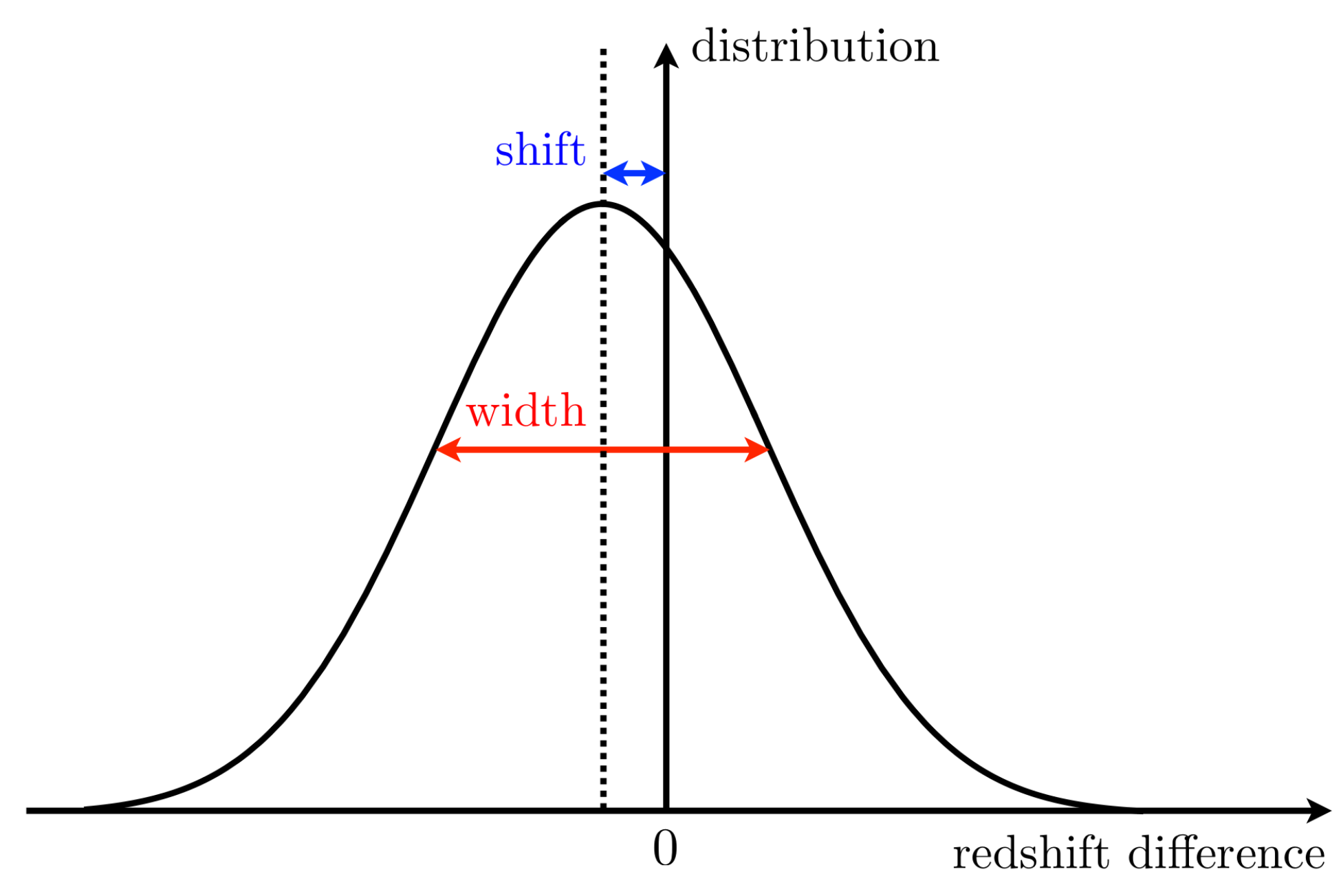}
 \caption{We can construct the distribution of the redshift difference for all members of the cluster situated at a fixed transverse separation from the BCG. The width of the distribution is governed by the linear Doppler term, while the shift (which is negative) is determined by gravitational redshift and by second-order Doppler contributions. Since the width is typically 100 times larger than the shift, it is necessary to have a very precise measurement of the shape (by stacking a larger number of clusters) to precisely determine the shift. } 
 \label{fig:distribution} 
 \end{figure}

The main challenge in measuring gravitational redshift in clusters resides in disentangling its effect from the Doppler shift due to the motions of galaxies, which are significantly larger. A powerful method, first proposed by~\cite{Nottale:1990, Cappi:1995, Kim:2004tc}, consists in comparing the observed spectroscopic redshift of the cluster centre with that of the member galaxies. As the central galaxy lies close to the bottom of the cluster potential, its gravitational redshift is larger than that of the other cluster members, leading to a net negative redshift difference between all members of the cluster and the central galaxy. On the other hand, the Doppler shift experienced by the cluster members with respect to the centre has no definite sign: since galaxies move randomly in a virialised cluster (with a dispersion determined by the mass of the cluster), the shift can be positive or negative, with mean value equal to zero.
The idea of~\cite{Nottale:1990, Cappi:1995, Kim:2004tc} was to exploit this intrinsically different behaviour of gravitational redshift with respect to Doppler shift to isolate the former. In practice, this can be achieved by constructing a histogram of redshift differences in a sample of stacked clusters. Gravitational redshift induces a shift of the mean of the distribution from zero,
 while the width of the distribution is mainly controlled by the Doppler effect, see figure~\ref{fig:distribution}.
By stacking a sufficient number of clusters, the distribution can be reconstructed with enough precision to allow for a detection of the typical shift of about $- 10$\,km/s induced by gravitational redshift.\footnote{The redshift (and hence the redshift difference) is dimensionless, since it is defined as a ratio of frequencies. Measurements are however often quoted in km/s, multiplying the redshift by the speed of light $c$.} The first measurement of this kind was performed in~\cite{Wojtak:2011ia} with data from the Sloan Digital Sky Survey (SDSS), followed by updated results with subsequent data releases \cite{Sadeh:2014rya, Jimeno:2014xma, eBOSS:2021ofn, Rosselli:2022qoz}.

The signal extracted in this way, however, requires a careful modelling and interpretation. As pointed out in~\cite{Zhao:2012gxk, Kaiser:2013ipa, Cai:2016ors}, gravitational redshift is not the only effect causing a shift of the mean of the distribution from zero, but there are several additional corrections. More precisely, ref.~\cite{Zhao:2012gxk} showed that the transverse Doppler effect due to the transverse motion of galaxies with respect to the line of sight contributes to the shift; ref.~\cite{Kaiser:2013ipa} added contributions arising from the impact of the galactic motions on the observed flux and from the fact that we observe galaxies along our past light cone; and ref.~\cite{Cai:2016ors} derived additional corrections due to the fact that both the expansion of the Universe and the velocity of galaxies inside the cluster change between the emission time of the central galaxy and that of the cluster members. These different contributions, however, were identified at different stages on the basis of kinematic considerations and a coherent scheme to calculate all the relevant terms is still missing.

In this work, we perform the first relativistic derivation of the gravitational redshift signal from galaxy clusters including all corrections within a consistent framework. We employ the tools developed within perturbation theory applied to galaxy clustering \cite{Yoo:2009au, Yoo:2010ni, Challinor:2011bk, Bonvin:2011bg,Jeong:2011as,Yoo:2014sfa, Bertacca:2014dra, DiDio:2014lka,DiDio:2016ykq,DiDio:2020jvo} and construct a coherent formalism to model the scales involved in this measurement. We demonstrate that one of the effect introduced in~\cite{Kaiser:2013ipa}, the so-called light-cone effect, does not contribute to the signal when the density profile is evaluated on a constant time hypersurface. Our formalism allows for a detailed comparison of the amplitude of gravitational redshift with respect to the other contributions (which act as a contamination). We find that the contamination of the gravitational redshift signal is of order unity for galaxy members close to the central galaxy, at distances up to $\sim 1.5\,{\rm Mpc}$, and decreases to approximately $20\% - 30\%$ at $\sim 6\,{\rm Mpc}$, depending on galaxy population properties such as magnification bias and spectral index.

Our formalism has the advantage of being fully model-independent: it does not rely on a specific theory of gravity, nor on the validity of the weak equivalence principle, which is encoded in Euler equation. This is of particular importance, since gravitational redshift is directly sensitive to the temporal distortions in the metric, which enter the Euler equation. Hence, gravitational redshift provides a natural observable to test the weak equivalence principle on scales of a few megaparsecs. This well complements other tests of the weak equivalence principle at different scales, like the dipole modulation of the two-point correlation function of galaxies or its power spectrum, see e.g.~\cite{Bonvin:2018ckp,Bonvin:2020cxp,Umeh:2020cag,Bonvin:2022tii,Castello:2024jmq,Castello:2024lhl}, the consistency relation between $n$ and $(n+1)$-point correlation functions~\cite{Kehagias:2013rpa,Creminelli:2013nua}, or the morphology of galaxies~\cite{Kesden:2006zb,Desmond:2020gzn}.

The rest of the manuscript is structured as follows: in section~\ref{sec:model}, we calculate the redshift difference between the central galaxy of a cluster and its members up to second order in the weak-field expansion, which is the appropriate perturbation scheme on the scale of clusters. In section~\ref{sec:stack}, we then derive the signal stacked over many clusters and in section~\ref{sec:BCG_velocity}, we calculate how the gravitational redshift signal changes if the central galaxy is not exactly at the bottom of the cluster's gravitational potential. In section~\ref{sec:numerical}, we numerically compute the amplitude of the different contributions. We then compare our expression with previous literature in section~\ref{sec:compare} and conclude in section~\ref{sec:conclusion}. The appendices contain additional steps in the derivations and plots.

\section{Modelling of the gravitational redshift observable}
\label{sec:model}

\subsection{Weak-field expansion}
We assume that the Universe is described by a perturbed flat Friedmann-Lema\^itre-Robertson-Walker (FLRW) metric. When describing an observable, we have the freedom to work in any gauge. In this paper, we choose the conformal Newtonian gauge, where the FLRW line element takes the form
\begin{eqnarray} \label{eq:metric}
    ds^2 = a(\eta) \left[ -\left( 1+2 \Psi \right)d\eta^2 +\left( 1-2 \Phi \right)\left( d\chi^2 + \chi^2 d\Omega^2 \right) \right]\, .
\end{eqnarray}
Here, $a(\eta)$ is the scale factor, $\eta$ denotes the conformal time, $\chi$ is the comoving radial coordinate and $\Phi$ and $\Psi$ are the two Bardeen gravitational potentials. 
We work in natural units where the speed of light is set to 1.

We aim to model redshift differences for sources separated by distances smaller than the typical cluster scales, i.e.~a few megaparsecs. Since density fluctuations can be large on these scales, standard perturbation theory is insufficient, and we must treat density fluctuations  $\delta=\delta\rho/\bar \rho$ non-perturbatively (here, a bar denotes quantities in the background). At the same time, gravitational potentials are small in this regime, typically of the order of $10^{-5}$, approaching unity only in the proximity of black holes. Hence, we can treat them linearly. The peculiar velocities are proportional to the gradient of the gravitational potential, and in clusters they are typically of the order $v\sim 10^3 \ {\rm km/s }$, corresponding to $v\sim 3 \times 10 ^{-3}$ in natural units. This leads to $\Psi \sim v^2$, as is also predicted by the virial theorem. 
Therefore, we need to account for the velocities up to second order. More formally, we can relate the velocities and gravitational potentials to the density perturbations through the continuity and Poisson equations, respectively, leading to
\begin{equation}
   \delta \sim \left( k/\HH \right) v \sim \left( k /\HH \right)^2 \Psi \, ,
\end{equation}
where $k$ is the Fourier mode and $\HH$ is the conformal Hubble parameter. We therefore define the weak-field expansion parameter $\epsilon_\HH \equiv \HH/k$ and consistently consider all terms up to $\epsilon_\HH^2$. This means that we keep $\delta$ at all orders, $v\sim \eH \delta$ up to second order and $\Psi\sim \eH^2\delta$ at linear order.\footnote{In real space, the weak-field parameter $\epsilon_\HH$ is given through the inverse of the Laplacian operator. However, for the sake of simplicity, we will use $\epsilon_\HH$ both in real and Fourier space.} Note that taking the gradient of a field enhances its amplitude by $\eH^{-1}$. 
As discussed in section~\ref{sec:compare}, third-order perturbations $\mathcal{O}(\epsilon_\HH^3)$ vanish due to symmetry reasons, implying that our derivation is valid up to this order.

\subsection{Redshift difference}
\label{subsection:redshift_difference}

 \begin{figure}[t!]
 \centering
	\includegraphics[width=1\columnwidth]{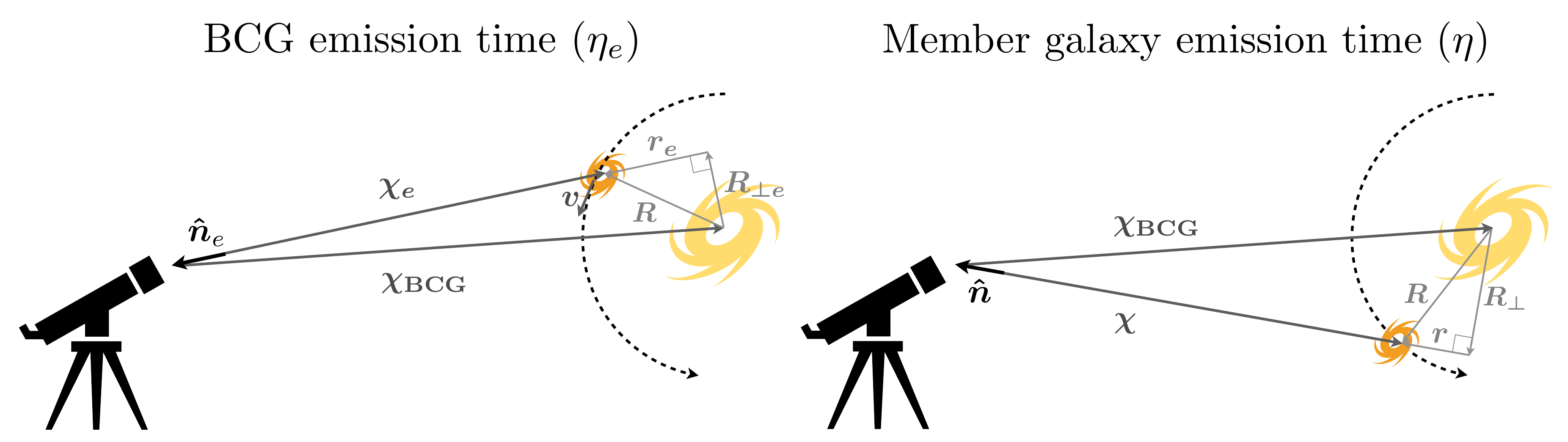}
 \caption{
 Geometrical setup with the relevant quantities defined in a cluster at the emission time of the BCG (left panel) and a generic member galaxy (right panel). Note that we have defined $\boldsymbol{r} \equiv (\boldsymbol{R}\cdot \hat{\boldsymbol{\chi}}) \, \hat{\boldsymbol{\chi}}$.  The motion of the member galaxy has been exaggerated for illustrative purposes. 
 In both panels, $\boldsymbol{\chi}_{\rm BCG}$ denotes the comoving distance to the BCG at its emission time. For the quantities related to the member galaxy, we distinguish between their values at the emission time $\eta$ of the member galaxy (without any subscript) and at the emission time $\eta_e$ of the BCG (with the subscript $e$).
 }
 \label{fig:geometrical_setup} 
 \end{figure}
 
The geometrical setup for the calculation of the redshift difference is shown in figure~\ref{fig:geometrical_setup}. In each cluster, we take the Bright Central Galaxy (BCG)  as a proxy for the centre and use it as the reference to calculate the redshift difference:
\begin{equation} \label{eq:def_Deltaz}
    \Delta z = \frac{z-z_c}{1+z_c}= \frac{1+ z}{1+z_c}-1\, ,
\end{equation}
where $z$ and $z_c$ denote the observed redshifts of a given member galaxy and of the BCG, respectively. The normalisation $1/(1+z_c)$ properly rescales the redshift difference by the scale factor when stacking clusters at different redshifts. 

The redshift of both the BCG and the galaxy member can be expressed in terms of a background redshift and a perturbation: $z= \bar z + \delta z$. Keeping terms up to order $\eH^2$, the perturbation is given by (see e.g.~\cite{Breton:2018wzk,DiDio:2020jvo})
\begin{equation}
\label{eq:deltaz}
     \delta z = \left(- \ndv - \Psi +  \frac{v^2}{2}\right) \left( 1 + \bar z \right) + \mathcal{O} \left( \epsilon_\HH^3 \right)\,.
\end{equation}
Here, $v$ is the norm of $\bv$ and $\ndv\equiv \boldsymbol{v}\cdot\bn$, with $\bn$ the unit vector pointing from the source to the observer. The first term in eq.~\eqref{eq:deltaz} corresponds to the linear Doppler contribution, the second one encodes gravitational redshift and the third one is the transverse Doppler term.\footnote{Strictly speaking, the transverse Doppler term refers to the contribution proportional to the transverse velocity $v_\perp^2$. In previous literature, however, the full second-order Doppler contribution proportional to $v^2$ has been called transverse Doppler, see e.g.~\cite{Kaiser:2013ipa}. We adopt this terminology here.} 
In eq.~\eqref{eq:deltaz}, we have neglected the integrated Sachs-Wolfe contribution, which depends on the time variation of the potentials and hence is subdominant with respect to gravitational redshift. 

We can now perform an expansion of the redshift difference $\Delta z$ in eq.~\eqref{eq:def_Deltaz}. Since the linear Doppler contribution to $\delta z$ in eq.~\eqref{eq:deltaz} is of order $\eH$, we need to expand $\Delta z$ up to second order in $\delta z$ and $\delta z_c$ to keep all velocity corrections up to $\mathcal{O}\left(\eH^2 \right)$,
\begin{eqnarray} \label{eq:2.5}
        \Delta z  &=&
        \frac{\bar z - \bar z_c}{1+ \bar z_c}  \left[ 1 
    +\frac{\delta z}{\bar z - \bar z_c} - \frac{(1+ \bar z)\delta z_c}{\left( \bar z - \bar z_c\right) \left( 1 + \bar z_c \right)} 
    + \frac{\delta z_c}{\left( \bar z - \bar z_c \right) \left( 1+ \bar z_c \right)} \left( \frac{1+ \bar z}{1 +\bar z_c}\delta z_c - \delta z \right)
    \right]\, 
    \nonumber \\
    &&
    + \,
    \mathcal{O}\left( \epsilon_\HH^3 \right) \, .
\end{eqnarray}
We notice that in addition to the terms proportional to $\delta z$ and $\delta z_c$, there is a term proportional to the difference of the background redshifts $\bar{z}-\bar{z}_c$. This term arises from the fact that a member galaxy and a BCG detected at the same time by the observer emitted their light at different times, since they are not at the same comoving distance from the observer. During this time interval, the background redshift of the cluster varies due to the expansion of the Universe, leading to a non-zero $\bar{z}-\bar{z}_c$, as was first shown by~\cite{Cai:2016ors}. To compute this term, we denote the emission time of the member galaxy with $\eta$ and that of the BCG with $\eta_e$, using the same subscript $e$ for functions that are evaluated at $\eta_e$ in the following. By considering that $\Delta \eta \equiv \eta- \eta_e$ is much smaller than the Hubble time $\HH^{-1}$, we expand $\bar{z}$ around $\bar{z}_c$ in powers of $\HH \Delta \eta$. Since $\Delta \eta$ is of the order of the typical pair separation in the cluster, we have $\HH \Delta \eta  \sim \frac{\HH}{k} =\epsilon_\HH$. Hence, going to second order in the weak-field expansion implies considering second-order perturbations in $\bar{z}$ around $\eta_e$: 
\begin{eqnarray} \label{eq:2.6}
 1+ \bar z &=& \frac{1}{a(\eta)} = \left( 1+\bar z_c\right) \left[ 1 - \HH_e  \Delta \eta + \left( \HH_e^2 - \frac{\ddot a_e}{2 a_e} \right) \left( \Delta \eta \right)^2 \right] {+ \, \mathcal{O} \left( \epsilon_\HH^3 \right)}
   \nonumber \\
   &=&
   \left( 1+\bar z_c\right) \left[ 1 - \HH_e \Delta \eta + \frac{1}{2} \left( \HH_e^2 - \dot\HH_e \right) \left( \Delta \eta \right)^2 \right]
   {+ \, \mathcal{O} \left( \epsilon_\HH^3 \right)}
   \, ,
\end{eqnarray}
where a dot denotes a derivative with respect to conformal time $\eta$. Thus, we obtain
\begin{eqnarray}
\label{eq:barz_te}
            \frac{ \bar z - \bar z_c}{1 + \bar z_c }&=& 
            - \HH_e \Delta \eta + \frac{1}{2} \left(1 - \frac{\dot\HH_e}{ \HH_e^2}\right)\left( \HH_e \Delta \eta \right)^2 {+ \, \mathcal{O} \left( \epsilon_\HH^3 \right)} \, .
\end{eqnarray}
We also expand $\delta z$ around $\eta_e$, only up to first order in this case, as $\delta z$ is already a first-order quantity on its own,
\begin{equation}
\label{eq:deltaz_te}
    \delta z = \left( \delta z \right)_{\eta_e} + \partial_\eta \left( \delta z \right)_{\eta_e} \Delta \eta   {+ \, \mathcal{O} \left( \epsilon_\HH^3 \right)} \, .
\end{equation}  

We can now relate the time difference $\Delta \eta$ to the line-of-sight separation between a given member galaxy and the BCG, which we denote by $r$. Along the light cone, we simply have
\begin{equation}
\label{eq:t_r}
    \Delta \eta =-\boldsymbol{R}\cdot \hat{\boldsymbol{\chi}} =-   r \, ,
\end{equation}
where $\boldsymbol{R}$ is the vector connecting the BCG and the member galaxy, see figure~\ref{fig:geometrical_setup}.
We aim to express the final redshift difference on constant-time hypersurfaces, where cluster properties can be modelled and directly tested with N-body simulations, see e.g.~\cite{Cai:2016ors}. Since the member galaxy is moving, $r$ is a function of time and denotes the separation when the member emits the signal observed at a given time $\eta_{\rm obs}$ (right panel of figure~\ref{fig:geometrical_setup}).\footnote{The line-of-sight distance $r$ increases for member galaxies that are further away than the BCG with respect to the observer, as shown in figure~\ref{fig:geometrical_setup}. Indeed, we have $\Delta \eta<0$ and $r>0$ for $\eta<\eta_e$.} In the following, we instead adopt $\eta_e$ as the reference time. Hence, we aim to express eq.~\eqref{eq:t_r} in terms of $r_e$, i.e.~the distance when the BCG emits the light detected at $\eta_{\rm obs}$ (left panel of figure~\ref{fig:geometrical_setup}).  
By performing a Taylor expansion of $r(\eta)$ around $\eta_e$, we obtain
\begin{equation} \label{eq:r_expanded_Deltaeta}
    r= r_e +  \dot r_e \Delta \eta \, + \, \mathcal{O} \left( \epsilon_\HH^3 \right)\, ,
\end{equation}
where we only consider terms up to $\mathcal{O} \left(\Delta \eta \right)$ as $r$ is already a first-order quantity following from eq.~\eqref{eq:t_r}. This leads to
\begin{equation}
\label{eq:Delta_t}
 \Delta \eta = - \frac{r_e}{1+\dot r_e } \, + \, \mathcal{O} \left( \epsilon_\HH^3 \right)  = - r_e + r_e \dot r_e   {+ \, \mathcal{O} \left( \epsilon_\HH^3 \right)} \, .
\end{equation}
Here, $\dot r_e$ is the velocity of the member in the frame at rest with respect to the observer, whose origin is set by the BCG at its emission time $\eta_e$. This leads to $\dot r_e =- \ndv$, where the negative sign is due to the direction of the vector $\bn$, chosen to point from the source to the observer (see figure~\ref{fig:geometrical_setup}).

Inserting eq.~\eqref{eq:Delta_t} into eqs.~\eqref{eq:deltaz_te} and~\eqref{eq:barz_te} and using eq.~\eqref{eq:deltaz}, we obtain 
\begin{eqnarray} 
    \Delta z &=& \HH_e r_e + \frac{1}{2}\left( 1- \frac{\dot \HH_e}{\HH_e^2} \right) \HH_e^2 r_e^2 
   + \frac{\left(\delta z\right)_{\eta_e} - \delta z_c}{1+\bar z_c} \left( 1 - \frac{\delta z_c}{1+ \bar z_c}\right)
   \nonumber \\
   &&
       - r_e  \HH_e  \left( \dot r_e + \frac{\delta z_c}{1+\bar z_c} \right)
    - r_e \frac{\partial_\eta \left( \delta z \right)_{\eta_e}}{1+\bar z_c}  {+ \, \mathcal{O} \left( \epsilon_\HH^3 \right)} 
     \nonumber \\
    &=&    \HH_e r_e - \Delta \ndv + \frac{1}{2}\left( 1- \frac{\dot \HH_e}{\HH_e^2} \right) \HH_e^2 r_e^2 
    + \HH_e r_e \ndv
    + r_e \dot \ndv 
+ \frac{\Delta v^2}{2} - \Delta \ndv^2
- \Delta \Psi
\nonumber \\
&&
- \HH_e r_e \Delta \ndv
+ \ndv \Delta \ndv
 {+ \, \mathcal{O} \left( \epsilon_\HH^3 \right)} 
\, , \label{eq:Deltaz_inter}
\end{eqnarray}
where all the quantities are now evaluated at $\eta=\eta_e$  
and we have used 
\begin{equation}
    \partial_\eta \left( \delta z \right)_{\eta_e}= - \left( 1 + \bar z_c \right) \left[  \dot \ndv - \HH \ndv \right]_{\eta_e}  + \mathcal{O} \left( \epsilon_\HH^2 \right) \, .
\end{equation}
Eq.~\eqref{eq:Deltaz_inter} depends on the gravitational potential difference between the member galaxy and the BCG at time $\eta_e$, $\Delta\Psi\equiv\Psi(r_e)-\Psi_c$, on the radial velocity difference $\Delta \ndv\equiv v_{\parallel}(r_e)-v_{\parallel\, c}$ and on the quadratic difference $\Delta v^2\equiv v^2(r_e)-v^2_{c}$.
Using 
\begin{equation}
\ndv\Delta\ndv=\frac{1}{2}\Delta\ndv^2+\frac{1}{2}(\Delta\ndv)^2\, ,    
\end{equation}
we can rewrite $\Delta z$ as 
\begin{eqnarray}
    \Delta z &=& - \Delta \Psi+ \HH_e r_e + \frac{1}{2}\left( 1- \frac{\dot \HH_e}{\HH_e^2} \right) \HH_e^2 r_e^2 - \Delta \ndv +\frac{1}{2}\Delta v^2 + \frac{1}{2}\left( \Delta \ndv \right)^2 - \frac{1}{2}\Delta \ndv^2 
    \nonumber \\
    &&
    + r_e \dot  \ndv + \HH_e r_e\ndv - \HH_e r_e \Delta \ndv + \mathcal{O} \left( \epsilon_\HH^3 \right)
    \, . \label{eq:Deltaz_final}
\end{eqnarray}
We see that besides the gravitational redshift contribution, $-\Delta\Psi$, there are various effects that impact the redshift difference: the second and third term in the first line encode the evolution of the background between the emission time of the member and the BCG; the fourth term is the linear Doppler contribution, while the last three terms in the first line are the second-order Doppler contributions (which include the transverse Doppler effect). Lastly, the terms in the second line contain the impact of the background cosmological evolution in combination with linear velocity corrections on the velocity.
The original idea of~\cite{Nottale:1990, Cappi:1995, Kim:2004tc} was to exploit the different symmetries of these contributions inside a cluster to isolate the gravitational redshift contribution, $-\Delta\Psi$, and reduce the contamination by cancelling the dominant linear Doppler contribution, $-\Delta \ndv$. In section \ref{sec:galaxy_weights}, we will discuss how this can be put into practice for a catalogue of galaxy clusters.

Eq.~\eqref{eq:Deltaz_final} has been derived in a fully model-independent way, only assuming that light travels on null geodesics of the metric in eq.~\eqref{eq:metric}. To compare with previous results derived in~\cite{Cai:2016ors}, we need to further assume the validity of the Euler equation,
\begin{eqnarray}
\label{eq:Euler}
    \dot \ndv + \HH \ndv - \partial_\chi \Psi  + \bv \cdot \nabla \ndv=0 \, .
\end{eqnarray}
Inserting~\eqref{eq:Euler} in~\eqref{eq:Deltaz_final}, we obtain 
\begin{eqnarray} \label{eq:DeltaZ_Euler}
     \Delta z_{\rm Euler}     &=&  \HH_e r_e + \frac{1}{2}\left( 1- \frac{\dot \HH_e}{\HH_e^2} \right) \HH_e^2 r_e^2 - \Delta \ndv +\frac{\Delta v^2}{2} + \frac{\left( \Delta \ndv \right)^2}{2} - \frac{\Delta \ndv^2}{2} - \Delta \Psi
    \nonumber \\
    &&
    + r_e \partial_{r_e} \Psi - \HH_e r_e \Delta \ndv  - r_e \bv \cdot \nabla \ndv
     {+ \, \mathcal{O} \left( \epsilon_\HH^3 \right)} 
    \, ,
\end{eqnarray}
where we have replaced the derivative of the gravitational potential with respect to the comoving distance, $\partial_\chi \Psi$, with the derivative with respect to the coordinate $r_e$, $\partial_{r_e} \Psi$. Indeed, since the term $r_e \partial_\chi \Psi$ is of second order in the weak-field expansion, we can treat it in the flat-sky approximation, where the line-of-sight to the all galaxies in the cluster are taken to be parallel, and hence the two derivatives are equivalent. Eq.~\eqref{eq:DeltaZ_Euler} agrees with eq.~(17) of~\cite{Cai:2016ors}, apart from the last term that is neglected there. This term should however be included, as it is of the same order as the others in the weak-field approximation due to the gradient of $\ndv$. As shown in section~\ref{sec:compare}, excluding this term results in a difference of approximately $12\% $ in the final observable.

\subsection{Average over all cluster members: weighting by the galaxy distribution}\label{sec:galaxy_weights}

\begin{figure}[t!]
 \centering
\includegraphics[width=0.7\columnwidth]{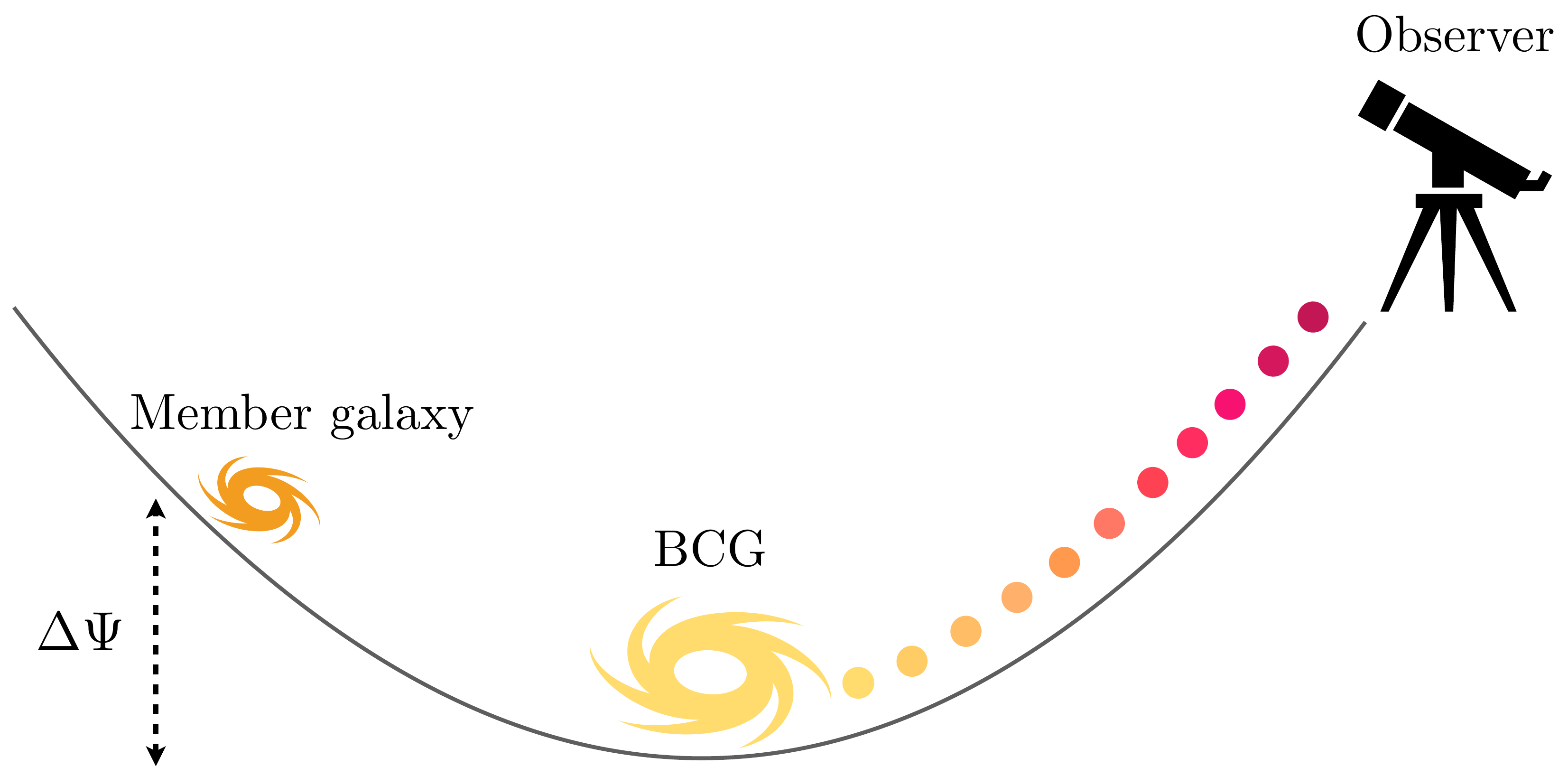}
 \caption{Sketch of the gravitational redshift effect. The photon emitted by a galaxy is redshifted when escaping a gravitational potential to reach the observer. In order to extract this signal, we take the observed redshift difference between each member galaxy in the cluster and the BCG, which we consider as a proxy for the bottom of the cluster gravitational potential. On average, member galaxies are located higher in the potential, thus they are less gravitationally redshifted than the BCG. This yields a net measured blueshift with this method.
 \label{fig:sketch}}
 \end{figure}

The redshift difference $\Delta z$ in eq.~\eqref{eq:Deltaz_final} is dominated by the linear Doppler contribution, $\Delta\ndv$. Nevertheless, inside a cluster, this term obeys different symmetries than the gravitational redshift contribution, $\Delta\Psi$, which allow us to isolate the latter. Since a cluster is a virialised object, its velocity distribution is centred around zero and symmetrical, with on average the same number of galaxies moving away and towards the observer. Hence, if we average the redshift difference between the BCG and all the members situated at a fixed distance from it, the resulting redshift distribution will also be centred around zero and symmetrical. Contrary to $\Delta\ndv$, the gravitational redshift contribution has a definite sign. Since member galaxies are located higher up in the gravitational potential of the cluster than the BCG (which we take as a proxy for the bottom), the redshift difference induced by gravitational redshift is always negative: all members are blueshifted with respect to the BCG. Hence, when performing the average over all members at fixed distance from the centre, we obtain a net blueshift of the distribution. By measuring this shift, we can therefore isolate $\Delta\Psi$ from the linear Doppler contribution, as shown in figure~\ref{fig:sketch}.

\begin{figure}[t!]
 \centering
\includegraphics[width=1.0\columnwidth]{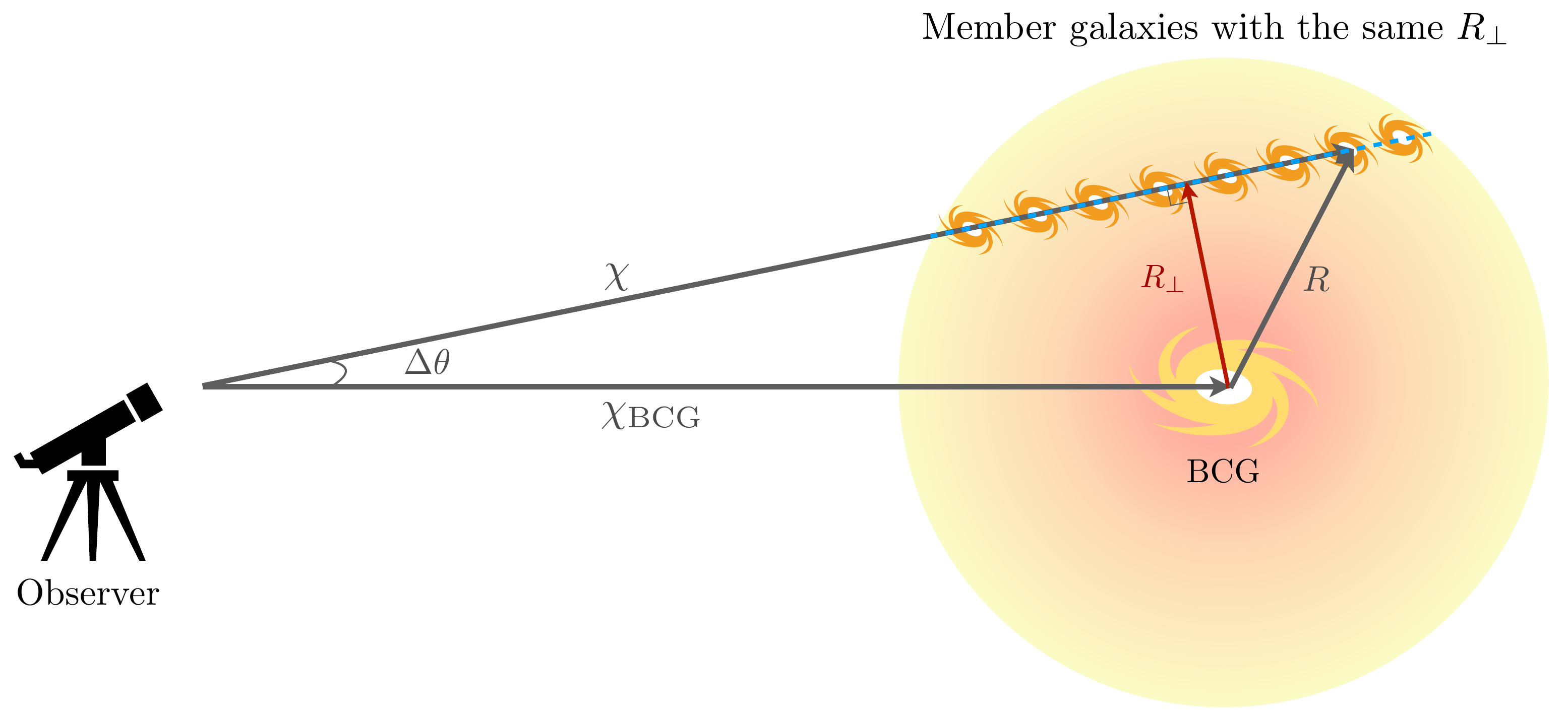} 
 \caption{\label{fig:R_perp}
The quantity $\Delta z$ is averaged over all galaxy members situated at the same $R_\perp$. While $R_\perp$ is uniquely defined in the flat sky, there are different possible definitions in the full sky. Here, we define $R_\perp = \sin(\Delta \theta) \chi_\mathrm{BCG}=\Delta\theta\,\chi_\mathrm{BCG}+\mathcal{O}(\epsilon^3_\HH)$, which means that we average over all member galaxies along the dashed blue segment.}
 \end{figure}

To model this blueshift, we need to weight the redshift difference in eq.~\eqref{eq:Deltaz_final} by the observed number of member galaxies in the cluster. Following the measurements of~\cite{Wojtak:2011ia}, we calculate the mean redshift difference averaged over galaxies situated at fixed transverse distance $R_\perp$ from the BCG, see figure~\ref{fig:R_perp}: 
\begin{equation} \label{eq:observable_def}
\overline {\Delta z}(R_\perp) \equiv \frac{\int dz \  n_g \left( \bn ,z, F \ge F_* \right) \Delta z }{\int dz \ n_g \left( \bn, z, F \ge F_*  \right)}\Bigg{|}_{R_\perp} \,.
\end{equation}
Here, $n_g \left(  \bn,z, F \ge F_* \right) = dN/dz/d\Omega$ is the number of observed galaxies per redshift interval and solid angle that are brighter than the flux limit of the survey $F_*$. For simplicity, we will denote any dependence on $F \ge F_*$ with $F_*$ in the following. From eq.~\eqref{eq:observable_def}, we see that the mean redshift $\overline {\Delta z}(R_\perp)$ is not only affected by the redshift difference $\Delta z$, but also by perturbations to $n_g$, which is impacted by density fluctuations and by relativistic effects due to peculiar velocities and gravitational potentials~\cite{Yoo:2009au, Yoo:2010ni, Challinor:2011bk, Bonvin:2011bg,Jeong:2011as,Yoo:2014sfa, Bertacca:2014dra, DiDio:2014lka,DiDio:2020jvo}.

We note that the setup in eq.~\eqref{eq:observable_def} does not respect the isotropy of the clusters. Since clusters can be considered spherically symmetric to a good approximation, we would ideally average over all member galaxies at fixed distance from the centre (instead of fixed $R_\perp$), such that they would be located at the same height in the gravitational potential and hence share the same gravitational redshift difference with respect to the BCG. In practice, however, it is not possible to observationally identify member galaxies at fixed distance from the centre, since observations are made in redshift space. Therefore, the radial distance from the centre cannot be measured directly, but needs to be derived from the transverse distance and the redshift. Since the redshift is the observable of interest, it cannot be used at the same time to reconstruct the coordinates in the geometrical setup.
On the contrary, the reference quantity used in the measurements is the transverse distance $R_\perp$, which can be obtained for all member galaxies from the measured angular separation $\Delta \theta$ and the comoving distance to the BCG, i.e.~$R_\perp=\Delta\theta  \chi_{\rm BCG}$ (see figure~\ref{fig:R_perp}).

To account for the impact of $n_g$ on $\overline {\Delta z}(R_\perp)$ in eq.~\eqref{eq:observable_def},  we split it into a background contribution $\bar n_g (z, \bar{L}_*(z))$ and a perturbation $\Delta \left(  \bn,z, L_*(z,\bn) \right)$ (which we do not assume to be small): 
\begin{equation}
\label{eq:ng_def}
    n_g \left(  \bn,z, F_* \right) = \bar n_g (z,\bar L_*(z))   \big( 1+ \Delta \left(  \bn,z,L_*( \bn,z) \right)\big)\, .
\end{equation}
Here, $L_*(\bn,z)$ is the luminosity threshold related to the flux threshold as $F_*=\frac{L_*(\bn,z)}{4\pi D^2_L(\bn,z)}$, where $D_L$ is the luminosity distance and $\bar{L}_*(z)$ is the mean luminosity threshold averaged over the direction $\bn$. The dependence of $\Delta$ on $L_*(\bn,z)$ reflects the fact that galaxies observed at a fixed flux do not have the same intrinsic luminosity in an inhomogeneous universe, since the propagation of light (and hence the flux) is affected by inhomogeneities, see~\cite{Bonvin:2005ps,Hui:2005nm}.

We can now insert eq.~\eqref{eq:ng_def} into $\overline {\Delta z}(R_\perp)$ in eq.~\eqref{eq:observable_def} and evaluate the integral. 
In doing so, we remark that the integration over the redshift in eq.~\eqref{eq:observable_def} effectively corresponds to integrating along the line of sight to each member galaxy in the clusters, i.e.~across different times. Since cluster density profiles, as for instance the Navarro-Frenk-White (NFW) one~\cite{Navarro:1995iw}, are modelled in real space (i.e.~on a hypersurface of constant time), if we want to use such profiles, we need to change the integration variable to the background comoving distance, projected on a hypersurface of constant time. We take as a reference the BCG emission time $\eta_e$, as in section \ref{subsection:redshift_difference}, and aim to express the integral in terms of the distance $r_e$.
To do so, we first change the integration variable from the observed redshift $z$ to the comoving distance $\chi$, and we will later change from $\chi$ to $r_e$. We can write 
\begin{eqnarray}
\label{eq:change_chi}
    \int dz \ n_g  \left( \bn ,z, F_* \right) \Delta z 
    &=& \int d\chi \frac{dz}{d\chi} \bar n_g \left( z, \bar{L}_* \right)  \left( 1+ \Delta \left(  \bn,z, L_* \right) \right) \Delta z\, ,
\end{eqnarray}
where $dz/d\chi$ is the Jacobian arising from the change of coordinates, and $\bar n_g$ and $\Delta $ need to be expressed in terms of $\chi$. 

Since $\chi$ can be linked to the background redshift $\bar{z}$ with a one-to-one correspondence, we can equivalently express $\bar n_g$ and $\Delta $ in terms of $\bar{z}$. For $\bar{n}_g$, we have
\begin{equation}
\label{eq:barn}
\bar{n}_g(z,F_*)=\bar n_g \left( \bar z, \bar{L}_* \right)  \left( 1 +\frac{d \log \bar n_g(\bar{z},\bar{L}_*)}{d \bar z} \delta z \right)
 {+ \mathcal{O} \left( \epsilon_\HH^2 \right)} 
\, .
\end{equation}
The galaxy fluctuations $\Delta$ have been computed within perturbation theory at first order~\cite{Yoo:2009au,Yoo:2010ni,Bonvin:2011bg,Challinor:2011bk,Jeong:2011as,DiDio:2016ykq} and second order~\cite{Yoo:2014sfa,Bertacca:2014dra,DiDio:2014lka}, including all relativistic effects. These results, however, cannot be employed on cluster scales, where $\delta$ is of order unity and hence standard perturbation theory is not valid. We therefore need to compute $\Delta$ in the weak-field expansion, as we did for the redshift perturbation $\Delta z$ in eq.~\eqref{eq:Deltaz_final}. Since the lowest order contribution to $\Delta z$ is already at first order in the weak-field parameter $\epsilon_\HH$, we only need to consider $\Delta$ up to first order to compute $\overline {\Delta z}(R_\perp)$ in eq.~\eqref{eq:observable_def} at second order. A first-order expression for $\Delta$ in the weak-field expansion has already been derived in~\cite{DiDio:2020jvo} in terms of the observed redshift $z$. However, as for $\bar{n}_g$ in eq.~\eqref{eq:barn}, we need to express $\Delta$ in terms of the background redshift $\bar{z}$.\footnote{More precisely, we need the observed $\Delta$ at the observed redshift $z$, as this is the quantity entering the observed redshift difference, but expressed in terms of $\bar{z}$, which is the integration variable on the left-hand side of eq.~\eqref{eq:change_chi}.} The derivation is presented in
appendix~\ref{app:weakfield_numbercounts}, yielding the following result:
\begin{eqnarray}
\label{eq:Delta_resum}
   \Delta \left( \bn , z, L_* \right) 
&=& \left( 1+ \delta \right) \left( 1 + \frac{d \delta z}{d\bar z} \right)^{-1} \left( 1- \frac{d \log \bar n_g (\bar{z},\bar{L}_*)}{d \bar z} \delta z\right)
     \left[1 + \left( 1 - \frac{5}{2} s_b \left(3 +   \alpha \right) \right) \frac{\delta z}{1+\bar z}\right]
     \nonumber \\
     &&
 -1 + \mathcal{O} \left( \epsilon_\HH^2 \right) 
\, .
\end{eqnarray} 
All functions on the right-hand side are evaluated at the background redshift $\bar{z}$, with $s_b$ and $ \alpha$ denoting the magnification bias and the spectral index, respectively. Here, we have neglected the lensing effect due to the deflection field since, as we will explain in more detail at the end of the section, the asymmetry in this effect along the line of sight relative to the cluster centre is expected to be negligible~\cite{Bonvin:2013ogt,Breton:2018wzk}.

Inserting eqs.~\eqref{eq:barn} and~\eqref{eq:Delta_resum} into eq.~\eqref{eq:change_chi} and using 
\begin{equation}
\label{eq:dzdchi}
 \frac{dz}{d\chi}=\frac{d\bar{z}}{d\chi}\left( 1 + \frac{d \delta z}{d \bar z} \right)\, ,  
\end{equation}
we obtain 
\begin{eqnarray}
\label{eq:Delta_interm}
    \int \! dz\, n_g  \left(  \bn ,z , F_* \right) \Delta z 
       &=&\!
              \int\! d\chi  \frac{d \bar z}{d\chi} \bar n_g \left( \bar z, \bar{L}_* \right) 
 \left( 1+ \delta \right) 
     \left[1 + \left( 1 - \frac{5}{2} s_b \left(3 +   \alpha \right) \right) \frac{\delta z}{1+\bar z}\right]  \Delta z
      {+ \, \mathcal{O} \left( \epsilon_\HH^3 \right)} 
     \, .\nonumber\\
\end{eqnarray}
We can now write the number of galaxies $\bar n_g$ as the product of the density $\bar{\rho}_g$ and the volume per solid angle and redshift bin $\mathcal{\bar V}$, $\bar{n}_g = \bar{\rho}_g \,\bar{\mathcal{V}}$, and use
\begin{eqnarray}
\label{eq:jac}
 \frac{d \bar z}{d  \chi} &=&\frac{    \mathcal{\bar V} \left( \bar \chi \right)}{ \mathcal{\bar V} \left( \bar z \right)}  \, ,\\
    \frac{\delta z}{1+ \bar z} &= & - \ndv  {+ \mathcal{O} \left( \epsilon_\HH^2 \right)}  \, .
\end{eqnarray}
This yields
\begin{eqnarray}
\label{eq:2.18}
    \int dz \ n_g  \left(  \bn ,z , F_* \right) \Delta z 
       &=&
        \int d\chi  \rho_g^{\rm real} \big( \eta\left( \chi \right), \bn, \bar{L}_* \big) \mathcal{\bar V} \left( \chi \right)
 \left[1 + \left( -1 + \frac{5}{2} s_b \left(3 +   \alpha \right)  \right) \ndv \right]
        \Delta z \nonumber\\ 
        &&+ \mathcal{O} \left( \epsilon_\HH^3 \right)\, ,
\end{eqnarray}
where we have introduced the number density of galaxies in real space, 
\begin{equation} \label{eq:rho_real}
\rho^{\rm real}_g \left( \eta(\chi), \bn, \bar{L}_* \right) \equiv \bar \rho_g\left(\eta(\chi), \bar L_* \right)   \left( 1 + \delta\left(\eta(\chi),  \bn\right) \right)  \, .
\end{equation}

Before further changing variable from $\chi$ to $r_e$, it is worth considering the different terms in eq.~\eqref{eq:2.18}. We remark that the density weighting introduces three new contributions to $\overline {\Delta z}(R_\perp)$. First, we recognize the so-called light-cone term,$-\ndv$, which accounts for the fact that galaxies detected at a fixed time by the observer did not emit their signal on a surface of constant time in their reference frame. Hence, the density of galaxies as a function of $\chi$ (or $\bar{z}$) is not the same as the intrinsic one defined on their hypersurface of constant time.  Secondly, eq.~\eqref{eq:2.18} contains a term proportional to the magnification bias $s_b$, which accounts for the fact that the flux of galaxies is magnified (or de-magnified) due to their peculiar velocities, changing their number above the flux threshold. As discussed in appendix \ref{app:magnification_bias}, this generates corrections to the galaxy number count fluctuations that are proportional to the logarithmic derivative of the background source density, evaluated at the survey luminosity threshold $\bar{L}_*$. Lastly, we have a contribution proportional to the spectral index $\alpha$. This encodes the fact that if two galaxies located at fixed $\chi$ and emitting at the same frequency (or in the same frequency band) are moving with respect to each other, we do not observe the same part of their spectrum. As a result, their flux can be can magnified or de-magnified, again changing the number of observed galaxies. This term is not present when expressing $\Delta$ as a function of the observed redshift $z$ (as e.g.~\cite{Bonvin:2011bg}), since in this case, we always observe the same part of the spectrum, related to the observed frequency by $(1+z)$. We see that all the other contributions to $\Delta$ that are present at fixed $z$ are canceled through the integration in $\chi$. For example, the redshift-space distortions (RSD) term, which is present in eq.~\eqref{eq:Delta_resum} through the term $d\delta z/d\bar{z}$, cancels out with the Jacobian in eq.~\eqref{eq:dzdchi}. Physically, this is due to the fact that RSD are remapping the position of the galaxies within the cluster in redshift space, but this does not change the total number of galaxies that are observed. 

In eq.~\eqref{eq:rho_real}, we have explicitly written $\eta$ as a function of $\chi$, to emphasise that $\rho_g^{\rm real}$ is defined on the light cone. Hence, integrating over $\chi$ in eq.~\eqref{eq:2.18} corresponds to considering different times $\eta$. As discussed above, the distribution of galaxies can only be modelled at a fixed time, so the integration variable should be further changed from $\chi$ to the spatial distance $r_e$, defined at fixed time $\eta_e$. This is fully consistent, since at fixed $R_\perp$, $r_e$ completely determines the position of the member galaxy in the cluster. Expanding the zeroth-order terms $\mathcal{\bar V}(\chi)$ and $\rho^{\rm real}_g \left( \eta(\chi), \bn, \bar{L}_* \right)$ around $\eta_e$ at order $\eH$, we obtain 
\begin{eqnarray}
   && \hspace{-1.3cm} \int dz \ n_g  \left( \bn, z ,F_* \right) \Delta z 
    = \mathcal{\bar V}_e
\int dr_e \frac{d\chi}{dr_e} 
\rho^{\rm real}_g \left( \eta_e, r_e, \bar{L}_* \right) 
\nonumber \\
&& \qquad \times
\left[ 1 - \left( \frac{ {\mathcal{\dot{\bar V}}}_e}{\mathcal{\bar V}_e \HH_e} + \frac{\dot \rho_g^{\rm real}}{ \rho_g^{\rm real} \HH_e} \right) \HH_e r_e
+   \left(  \frac{5}{2} s_b \left(3 +   \alpha \right) -1 \right) \ndv   \right]
        \Delta z 
        + \mathcal{O} \left( \epsilon_\HH^3 \right)\, .
        \label{eq:Delta_re}
\end{eqnarray} 
The density profile of the cluster is now defined at fixed time $\eta_e$ in terms of intrinsic properties of the cluster: the intrinsic luminosity $\bar{L}_*$ and the distance to the centre $r_e$. Note that we do not need to expand the first-order contributions in $n_g$ around $\eta_e$, since they multiply $\Delta z$, which is already linear in $\epsilon_\HH$.

\begin{figure}[t!]
 \centering
	\includegraphics[width=1\columnwidth]{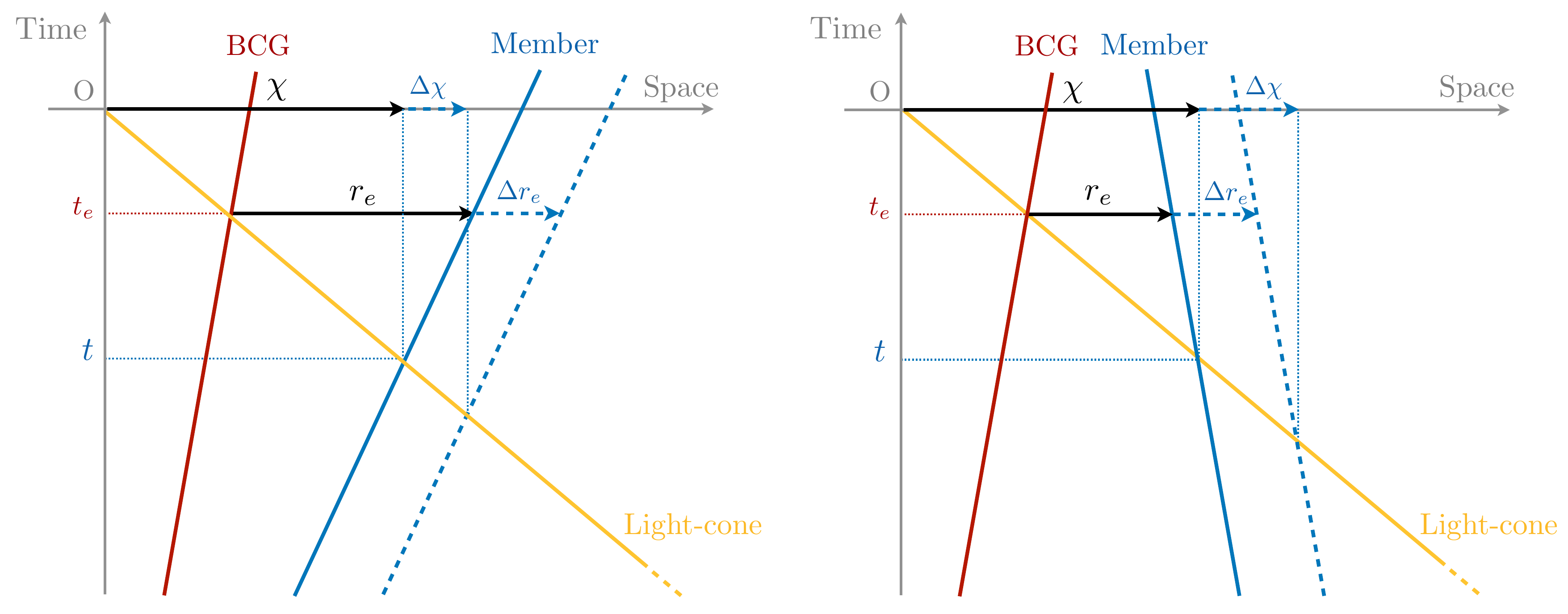}
 \caption{
 Illustration of the light-cone effect, showing the worldlines of the BCG (in red) and a member galaxy (in blue). On the left panel, the member galaxy is moving away from the BCG ($\dot{r}_e > 0$), leading to a change $\Delta r_e > \Delta \chi$, and hence $\frac{d \chi}{d r_e} > 1$. On the contrary, the member galaxy is moving towards the BCG on the right panel ($\dot{r}_e < 0$), leading to $\Delta r_e < \Delta \chi$, and hence $\frac{d \chi}{d r_e} < 1$.} 
 \label{fig:light-cone} 
 \end{figure}

We can now rewrite the Jacobian $d\chi/dr_e$ in eq.~\eqref{eq:Delta_re} according to the setup in figure~\ref{fig:geometrical_setup}. To do so, we first express $\chi$ in terms of $r$,
\begin{eqnarray}
\label{eq:chi_WA}
    \left\{
    \begin{array}{ll}
&\cos \left( \Delta \theta \right) \chi_{\rm BCG}= \chi- r \\
&\sin \left( \Delta \theta \right)\chi_{\rm BCG} = R_\perp
    \end{array}
    \right. \Rightarrow \chi = \chi_{\rm BCG} \left( 1 + \frac{r}{\chi_{\rm BCG}} - \frac{R_\perp^2}{2 \chi_{\rm BCG}^2} + \mathcal{O} \left( \epsilon^3_\HH  \right) \right) \, ,
\end{eqnarray}
where we have performed an expansion in $\Delta\theta$, keeping terms up to order $\epsilon^2_\HH$, i.e.~$\Delta \theta = \chi_{\rm BCG}/R_\perp + \mathcal{O}(\epsilon_\HH^3)$. This allows us to assess the impact of potential wide-angle corrections arising from the fact that the lines of sight to the BCG and the member galaxies are not parallel. Note that the wide-angle and the weak-field expansions are parametrically equivalent for comoving distances $\chi_{\rm BCG}$ of the order of the Hubble horizon $\HH_e^{-1}$, since $\left\{R_\perp, R, r \right\} \sim k^{-1}$. We then expand $r$ around $r_e$ according to eqs.~\eqref{eq:r_expanded_Deltaeta}--\eqref{eq:Delta_t} and compute the Jacobian keeping $R_\perp$ and the velocity $\bv$ fixed, to ensure that the coordinate mapping is consistent: 
\begin{eqnarray}
\label{eq:lightcone_chi_r}
      \left. \frac{d\chi}{dr_e}\right|_{(R_\perp,\bv)} 
     &=&
     \left. \frac{d}{d r_e} \right|_{(R_\perp,\bv)} \chi_{\rm BCG}  \left( 1 + \frac{r}{\chi_{\rm BCG}} - \frac{R_\perp^2}{ 2\chi_{\rm BCG}^2} + \mathcal{O} \left( \epsilon_\HH^3 \right) \right)
     \nonumber \\
     &=&  \left. \frac{dr}{d r_e} \right|_{(R_\perp,\bv)}  +\mathcal{O} \left( \epsilon_\HH^2 \right) 
       =\left. \frac{d}{d r_e} \right|_{(R_\perp,\bv)} \left( r_e - \dot r_e r_e \right) +\mathcal{O} \left( \epsilon_\HH^2 \right)
       \nonumber \\
    & =&  \left. \frac{d}{d r_e} \right|_{(R_\perp,\bv)} \left( r_e + \ndv r_e \right) +\mathcal{O} \left( \epsilon_\HH^2 \right)
     =1+ \ndv  +\mathcal{O} \left( \epsilon_\HH^2 \right)
  \, . \quad 
\end{eqnarray}
We remark that the final expression does not contain any wide-angle corrections,\footnote{This differs from the behavior of the dipole term in the cross-correlation of two galaxy populations, where wide-angle corrections cause the quadrupole to leak into the dipole already to first order~\cite{Bonvin:2013ogt,Bonvin:2014owa,Bonvin:2015kuc,Gaztanaga:2015jrs,Beutler:2018vpe,Beutler:2020evf}.
} as a consequence of defining the observable $\overline{\Delta z} \left( R_\perp \right)$ in terms of a fixed $R_\perp$.
The term $\ndv$ in the last line of eq.~\eqref{eq:lightcone_chi_r} is the light-cone effect, which relates a fixed comoving distance $\chi$ to a distance at fixed emission time $\eta_e$, as depicted in figure~\ref{fig:light-cone}. This term exactly cancels the light-cone effect in $\Delta$ that is present in eq.~\eqref{eq:2.18}.

Inserting eq.~\eqref{eq:lightcone_chi_r} into eq.~\eqref{eq:Delta_re}, we obtain
\begin{equation}
   \int dz \ n_g   \left( \bn, z ,F_* \right)  \Delta z =  \mathcal{\bar V}_e
\int dr_e 
\rho^{\rm real}_g \left( \eta_e, r_e, \bar{L}_* \right)  
\left[ 
 1- \mathcal{B} \HH_e r_e + \mathcal{R} \ndv 
\right] \Delta z 
+ \mathcal{O} \left( \epsilon_\HH^3 \right) \, , \label{eq:deltazv_final}
 \end{equation}   
where we have defined 
\begin{align}
\mathcal{B}&= \left( \frac{{\mathcal{\dot{\bar V}}}_e}{\mathcal{\bar V}_e \HH_e} + \frac{\dot{ \rho}_g^{\rm real}}{  \rho_g^{\rm real} \HH_e} 
\right)\, ,\\
\mathcal{R}&=\frac{5}{2} s_b \left(3 +   \alpha \right)\, . \label{eq:defR}
\end{align}
As discussed above, by expressing the integral and the galaxy density as functions of the spatial distance $r_e$ at the fixed time $\eta_e$, the light-cone term cancels out. This term arises when the galaxy density distribution is described in terms of a comoving distance spanning the light cone, while it does not appear when the density distribution is modelled on a hypersurface of constant time, as in our case. As a consequence, $\mathcal{R}$ only contains a term that is directly proportional to $s_b$ and vanishes if all galaxies in the cluster are well above the flux threshold of the survey of interest.

\subsection{Weighting with the velocity distribution}\label{sec:velocity_distribution}

The expression obtained in eq.~\eqref{eq:deltazv_final}, combined with eq.~\eqref{eq:Deltaz_final}, contains all the contributions to the mean redshift difference up to order $\epsilon_\HH^2$. This quantity depends on the gravitational potential difference, the velocity of the cluster members and also that of the BCG through eq.~\eqref{eq:Deltaz_final}. However, we cannot obtain a theoretical prediction for these velocities, since matter is virialised on cluster scales. Therefore, shell-crossing occurs and perturbation theory becomes insufficient to describe the internal motion of galaxies. Nevertheless, we can predict the velocity distribution of the cluster members, as this is linked to the mass of the cluster. Hence, we can compute the expectation value of the redshift difference by weighting it with this velocity distribution. 

To do so, we assume that the BCG is at the centre of the cluster (we will relax this assumption in section~\ref{sec:BCG_velocity}). In the non-relativistic limit, this implies that the difference $\boldsymbol{\Delta v}\equiv\boldsymbol{v}(r_e)-\boldsymbol{v}_c$ describes the velocity of the member galaxies in the rest frame of the cluster. We thus weight the redshift difference by the velocity distribution $f\left( \boldsymbol{\Delta v} \right)$ and integrate over $ \boldsymbol{\Delta v}$. For virialised clusters, $f(\boldsymbol{\Delta v})$ is centred on zero and even in $\boldsymbol{\Delta v}$, implying that all odd terms in $\boldsymbol{\Delta v}$ vanish in the integration and do not contribute to the observed redshift difference. We thus obtain 
\begin{eqnarray} \label{eq:2.22}
 && \hspace{-0.8cm}  \int d^3 (\Delta v)  f\left( \boldsymbol{\Delta v} \right) \int dz \ n_g  \left( \bn, z ,F_* \right)  \Delta z
        =\mathcal{\bar V}_e           \int d^3 (\Delta v)   f\left( \boldsymbol{\Delta v} \right)
   \nonumber \\
        &&         
    \times \int dr_e 
\rho^{\rm real}_g \left( \eta_e, r_e, \bar{L}_* \right)  \left[  r_e \HH_e  +r_e \dot \ndv_c + r_e \Delta \dot \ndv+ r_e \HH_e \left( 1+\mathcal{R} \right) {\ndv}_c + \frac{\Delta v^2}{2} - \frac{\Delta \ndv^2}{2} - \Delta \Psi 
    \right.  \nonumber \\
    &&  \left. \qquad \qquad
    + \left( \Delta \ndv \right)^2 \left( \frac{1}{2} - \mathcal{R} \right)
    +\frac{1}{2}r_e^2 \HH_e^2 \left( 1-2 \mathcal{B} - \frac{\dot \HH_e}{\HH_e^2}\right)
    \right] 
 + \mathcal{O}\left( \epsilon_\HH^3 \right) \, .
\end{eqnarray}
By integrating the square of the peculiar velocity over its velocity distribution, we obtain the velocity dispersion:  
\begin{eqnarray} \label{eq:vel_disp1}
    \int d^3 (\Delta v) f \left( \boldsymbol{\Delta v} \right) \Delta v^2 &=&   \int d^3 (\Delta v) f \left( \Delta v \right) \left( \Delta v \right)^2 = 3 \sigma_v^2 \, , \\
        \int d^3 (\Delta v) f \left( \boldsymbol{\Delta v} \right) \Delta \ndv^2 &=&   \int d^3 (\Delta v) f \left( \Delta v \right) \left( \Delta \ndv \right)^2 = \sigma_v^2 \, , \label{eq:vel_disp2} 
\end{eqnarray}
where $\sigma_v$ denotes the one-dimensional velocity dispersion. Here, we have assumed that the velocity distribution is isotropic and therefore only depends on the magnitude $\Delta v$, $f(\boldsymbol{\Delta v})=f(\Delta v)$, such that the 3-dimensional velocity dispersion in eq.~\eqref{eq:vel_disp1} equals 3 times the one-dimensional one in eq.~\eqref{eq:vel_disp2}. Moreover, under the assumption that the time evolution of the velocity distribution conserves its parity, we have
\begin{equation}\label{eq:vel_disp3}
     \int d^3 (\Delta v) f \left( \Delta v \right) \Delta \dot \ndv = \partial_\eta \left(     \int d^3 (\Delta v) f \left( \Delta v \right) \Delta  \ndv \right) -     \int d^3 (\Delta v) \partial_\eta f \left( \Delta v \right) \Delta  \ndv= 0 \, .
\end{equation}
With this, we obtain 
\begin{eqnarray}
 &&  \hspace{-2cm} \int d^3 (\Delta v)  f\left(\Delta v \right) \int dz \ n_g  \left( \bn, z ,F_* \right)  \Delta z
 \nonumber \\
 &=&  \mathcal{\bar V}_c 
\int dr_e 
\rho^{\rm real}_g \left( \eta_e, r_e , \bar{L}_*\right)  \Bigg[  r_e \HH_e  +r_e \dot \ndv_c + r_e \HH_e \left( 1+\mathcal{R} \right) {\ndv}_c  - \Delta \Psi 
  \nonumber \\
    &&  \qquad \qquad
    + \sigma_v^2 \left( \frac{3}{2} - \mathcal{R} \right)
    +\frac{1}{2}r_e^2 \HH_e^2 \left( 1-2 \mathcal{B} - \frac{\dot \HH_e}{\HH_e^2}\right)\Bigg]      
 + \mathcal{O}\left( \epsilon_\HH^3 \right) \, . \label{eq:z_intv}
 \end{eqnarray}

\subsection{Using the symmetries of the cluster}\label{sec:cluster_symmetries}

We can further simplify the integral over $r_{e}$ in eq.~\eqref{eq:z_intv} by employing the fact that the cluster density profile is symmetric along the line of sight:
\begin{equation}
   \rho^{\rm real}_g \left( \eta_e, r_e, \bar{L}_* \right) =\rho^{\rm real}_g \left( \eta_e, -r_e,\bar{L}_* \right) \, .
\end{equation}
This assumption may not be true for individual clusters, but once staking a sufficiently large number of clusters, we expect the resulting profile to be close to isotropic.\footnote{Ref.~\cite{Cai:2016ors} tested the assumption of spherical symmetry for individual clusters, finding large discrepancies for low-mass clusters. However, stacking a large number of clusters at all masses tends to isotropise the profile. For the sake of simplicity, here we work under the isotropic profile assumption.} Since the profile of the cluster sources the gravitational potential and its geometry drives the galaxy velocities, this symmetry argument also applies to $\Psi$ and $\sigma_v$:
\begin{eqnarray}
    \Psi \left( r_e \right) &=& \Psi \left( - r_e \right) \, , \\
        \sigma_v^2 \left( r_e \right) &=&      \sigma_v^2 \left( - r_e \right) \, .
\end{eqnarray}
By using these assumptions and adopting symmetric integral boundaries (in real space), the redshift difference simplifies to
\begin{eqnarray}\label{eq:Deltaz_all_weighted}
 && \hspace{-0.8cm}   
 \int d^3 (\Delta v)  f\left(\Delta v \right) \int dz \ n_g  \left( \bn, z ,F_* \right)  \Delta z
 \\
 &=&  \mathcal{\bar V}_c 
\int dr_e 
\rho^{\rm real}_g \left( \eta_e, r_e, \bar{L}_* \right)  \left[     - \Delta \Psi 
    + \sigma_v^2 \left( \frac{3}{2} - \mathcal{R} \right)
    +\frac{1}{2}r_e^2 \HH_e^2 \left( 1-2 \mathcal{B} - \frac{\dot \HH_e}{\HH_e^2}\right)\right]      
 + \mathcal{O}\left( \epsilon_\HH^3 \right) \, . \qquad\nonumber
 \end{eqnarray}
We remark that all linear terms in the weak-field expansion, i.e.~at~$\mathcal{O}\left( \epsilon_\HH \right)$, vanish. This is the fundamental symmetry argument allowing us to detect gravitational redshift: the linear Doppler contribution, which is orders of magnitude larger, gets canceled by exploiting the fact that it has no preferred sign and thus vanishes on average.\footnote{The same symmetry argument applies to the description of the large-scale structure in the Universe in the context of two-point correlations. Relativistic effects, including gravitational redshift, are subdominant compared to density and RSD, but they can be isolated by looking for a dipole in the two-point correlation function. This dipole is not affected by density and RSD, which only source even multipoles, see refs.~\cite{McDonald:2009ud,Bonvin:2013ogt,Bonvin:2014owa,Irsic:2015nla,Bonvin:2015kuc,Gaztanaga:2015jrs,Breton:2018wzk,DiDio:2018zmk,Beutler:2020evf,Bonvin:2023jjq}.}

The absence of linear contributions implies that we only need to compute the normalisation term in eq.~\eqref{eq:observable_def} at $0$th-order in the weak-field expansion,
\begin{eqnarray}\label{eq:normaliz}
 \int  dz \ n_g   \left( \bn, z ,F_* \right) 
 &=&\int dr_e \frac{d\bar z}{dr_e} \left( 1 + \frac{d\delta z}{d\bar z } \right) n_g   \left( \bn, z ,F_* \right)   +\mathcal{O} \left( \epsilon_\HH \right)
 \nonumber \\
 &=& \mathcal{\bar V}_e \int dr_e  \rho^{\rm real}_g \left( \eta_e, r_e, \bar{L}_* \right) +\mathcal{O} \left( \epsilon_\HH \right) \, .
\end{eqnarray}
Analogously to eq.~\eqref{eq:Delta_interm}, the RSD term, which is of $0$th-order in the weak-field expansion, cancels out between the Jacobian $\left( 1 + \frac{d\delta z}{d\bar z } \right)$ and the galaxy number counts $n_g   \left( \bn, z ,F_* \right) $ (see eq.~\eqref{eq:Delta_resum}).  Combining eqs.~\eqref{eq:Deltaz_all_weighted}--\eqref{eq:normaliz}, the expected redshift difference takes the final form
\begin{eqnarray}
\label{eq:Deltaz_obs}
        \langle \overline{\Delta z} \rangle_{R_\perp}  &=& \frac{\int d^3 (\Delta v)  f\left(\Delta v \right) \int dz \ n_g  \left( \bn, z ,F_* \right) \Delta z}{\int d^3 (\Delta v)  f\left(\Delta v \right) \int dz \ n_g   \left( \bn, z ,F_* \right)  } \\
        &=&\frac{\int dr_e 
\rho^{\rm real}_g \left( \eta_e, r_e, \bar{L}_* \right)  \left[     - \Delta \Psi 
    + \sigma_v^2 \left( \frac{3}{2} - \mathcal{R} \right)
    +\frac{1}{2}r_e^2 \HH_e^2 \left( 1-2 \mathcal{B} - \frac{\dot \HH_e}{\HH_e^2}\right)\right]  }{ \int dr_e  \rho^{\rm real}_g \left( \eta_e, r_e, \bar{L}_* \right) } + \mathcal{O} \left( \epsilon_\HH^3 \right) \, ,  \nonumber 
\end{eqnarray}
where the angle brackets on the left-hand side denote the average over the velocity distribution. In this final expression, there are only two types of contamination terms that remain once $\Delta z$ is integrated over the cluster members and the velocity distribution: second-order kinematic Doppler effects proportional to $\sigma^2_v$ and evolution effects proportional to $r_e^2\HH_e^2$.

As stated above, in our derivation we have neglected contributions from gravitational lensing. We now conclude this section by further motivating this choice. Since the lensing convergence
\begin{equation}
   \kappa_{\rm len} = \int_0^{\chi_*} d \chi \frac{\chi_*- \chi}{\chi_* \chi}  \frac{\Delta_{\Omega}(\Phi + \Psi)}{2} 
\end{equation}
is sourced by the angular Laplacian ($\Delta_\Omega$) of the metric perturbations, one might naively expect it to be of the order of the density fluctuations $\delta$, i.e.\ of order unity, suggesting a correction of the form $\rho_g^{\rm real} \to \rho_g^{\rm real}(1 + \kappa_{\rm len})$. However, this expectation does not hold in practice. While density perturbations can be large on small scales, the lensing signal is an integrated effect over a broad kernel, which smooths out small-scale inhomogeneities. As a result, we typically find $\kappa_{\rm len} \ll 1$, even in clusters. Moreover, lensing merely displaces structures on the sky and is averaged out once integrated over the whole sky by stacking multiple clusters. 
Even though lensing is not of order unity, one might still ask whether it contributes at perturbative level. In our derivation, such a contribution would appear in terms like $\kappa_{\rm len} v_\parallel$, if $\kappa_{\rm len}$ were treated as linear in $\epsilon_\mathcal{H}$. However, since the lensing convergence is uncorrelated to the velocity field, the term $\kappa_{\rm len} v_\parallel$ is set to $0$ by the integration over the velocity dispersion, as it contains an odd power in the velocity. Hence, lensing does not contribute to the signal at the order we are interested in.

\section{Stacking clusters}
\label{sec:stack}

\subsection{Expectation value of the stacked signal}
The typical magnitude of gravitational redshift, encoded in $\Delta\Psi$, is of the order of $c\, \delta z \sim 10 \ {\rm km/s}$ ($\sim 10^{-5}$ in natural units) \cite{Wojtak:2011ia}, while the velocity dispersion $\sigma_v$ is typically of the order of $1000 \ {\rm km/s}$ ($\sim 3\cdot 10^{-3}$ in natural units). The latter does not affect the shift in the redshift distribution, given by $\langle \overline{\Delta z} \rangle_{R_\perp}$ in eq.~\eqref{eq:Deltaz_obs}. However, the width of the distribution, given by the variance $\langle (\overline{\Delta z})^2 \rangle_{R_\perp}\rangle^{1/2}$, is directly proportional to $\sigma_v$. This implies that the shift is typically two orders of magnitude smaller than the width. Therefore, a precise measurement of the shift can only be achieved if the shape of the distribution (in particular its width) is well determined. This requires a large number of galaxy members. For independent measurements, the precision scales with the square root of the number of sources, requiring more than $10^4$ galaxies to obtain a measurement of the shape as precise as the amplitude of the shift. To achieve this, it is necessary to stack several clusters of galaxies, combining them into a single measurement of $\langle \overline{\Delta z} \rangle_{R_\perp}$. The stacked redshift difference is given by 
\begin{eqnarray} \label{eq:DeltaZ_stack}
    \langle \overline{\Delta z} \rangle_{R_\perp}^{\rm stack} &=& \frac{\int dM \frac{dN}{dM}\int dr_e 
\rho^{\rm real}_g \left( \eta_e, r_e, \bar{L}_* , M\right)  \left[     - \Delta \Psi 
    + \sigma_v^2 \left( \frac{3}{2} - \mathcal{R} \right)
    +\frac{1}{2}r_e^2 \HH_e^2 \left( 1-2 \mathcal{B} - \frac{\dot \HH_e}{\HH_e^2}\right)\right]  }{ \int dM \frac{dN}{dM}\int dr_e  \rho^{\rm real}_g \left( \eta_e, r_e, \bar{L}_* ,M\right) } 
    \nonumber \\
    &&
    + \mathcal{O} \left( \epsilon_\HH^3 \right)\, ,
\end{eqnarray}
where $\frac{dN}{dM}$ is the cluster mass function.\footnote{We remark that we could not have swapped the order of the operations in eq.~\eqref{eq:DeltaZ_stack}, performing the integral over the mass before the one over the velocity distribution from section \ref{sec:velocity_distribution}, as the velocity distribution depends on the cluster mass.} Eq.~\eqref{eq:DeltaZ_stack} is a key result of this paper, describing the expectation value of the observed redshift difference from stacked clusters, under the assumption that the BCG is at rest at the bottom of the gravitational potential. We see that the signal depends on the gravitational redshift difference weighted by the density profile, as well as on contaminations from kinematic effects at second order and evolution effects.

\subsection{Modelling of the signal using an NFW density profile} \label{sec:model_with_NFW}

To proceed further and numerically compute the expected signal, we need to adopt a functional form for the density profile of the clusters, $\rho^{\rm real}_g$. As discussed in section \ref{sec:cluster_symmetries}, we assume that each cluster is spherically symmetric, and we adopt the NFW density profile \cite{Navarro:1995iw} as done in previous analyses, see e.g.~\cite{Wojtak:2011ia, Rosselli:2022qoz},\footnote{
We have neglected the impact of the observer velocity in the galaxy number counts $n_g$ in eq.~\eqref{eq:Delta_interm}, and hence also in the cluster density profile. This would introduce a dipolar modulation in the gravitational redshift, which would anyway vanish by stacking a sufficient number of clusters. Moreover, we note that $\rho^{\rm real}_g \left(\eta_e, r_e, \bar{L}_* ,M \right)$ is a number density, while $\rho_{\rm NFW} \left( R \right)$ is a mass density. However, since $\rho^{\rm real}_g (\eta_e, r_e, \bar{L}_*, M)$ appears in both the numerator and denominator of eq.~\eqref{eq:DeltaZ_stack}, we can replace it with $\rho_{\rm NFW} (R)$.}
\begin{equation} \label{eq:NFW}
    \rho^{\rm real}_g \left(\eta_e, r_e, \bar{L}_* ,M \right)\propto \rho_{\rm NFW} \left( R \right) = \frac{R_s}{R}\frac{\rho_0}{\left( 1 + \frac{ R}{R_s} \right)^2}\, .
\end{equation}
Here, $R$ is the radial distance from the cluster centre,\footnote{We follow~\cite{Wojtak:2011ia} by adopting the comoving distance from the centre. Other references, see e.g.~\cite{Rosselli:2022qoz}, have chosen to express the distance in units of $r_{500}$. i.e.~the radius where the cluster density is $500$ times the critical density of the Universe. This allows them to take advantage of the cluster self-similarity in the stacking procedure.} related to $r_e$ and $R_\perp$ through $R^2=r_e^2+R_\perp^2$, see figure~\ref{fig:geometrical_setup}. The quantities $R_s$ and $\rho_0$ depend on the properties of individual halos and need to be constrained observationally. These parameters contain an implicit dependence on the luminosity threshold, as they characterise the observed distribution of galaxies in the cluster with luminosity above $\bar{L}_*$. Moreover, $R_s$ and $\rho_0$ are the only quantities in the NFW profile that evolve as a function of time and hence encode the full dependence on $\eta_e$. By stacking clusters at different redshifts, corresponding to different emission times $\eta_e$ of the respective BCGs, we are only sensitive to effective values of $R_s$ and $\rho_0$, describing the shape of the stacked cluster.

To determine $R_s$ and $\rho_0$, we need to assume an underlying theory of gravity, relating the geometry of spacetime to the distribution and motion of galaxies. For the purpose of our estimate of $\langle \overline{\Delta z} \rangle_{R_\perp}^{\rm stack}$, we assume the validity of general relativity, where the two metric potentials in eq.~\eqref{eq:metric} are equal: $\Psi = \Phi$. Moreover, the potentials are linked to the density profile through the Poisson equation, which for the NFW profile in eq.~\eqref{eq:NFW} takes the form
\begin{equation}\label{eq:Poisson_NFW}
    \Phi_{\rm NFW} \left( R \right)=\Psi_{\rm NFW} \left( R \right) = \frac{4 \pi G R_s^3 \rho_0}{R} \log\left( \frac{R_s}{R+R_s}\right) \, .
\end{equation}

From the position of galaxies in redshift space, we cannot directly reconstruct the density profile to obtain $R_s$ and $\rho_0$. However, we can reconstruct the mass distribution from the velocity dispersion $\sigma_v$, which can be measured from the width of the redshift distribution in a spectroscopic survey.
Again, this cannot be done in a fully model-independent way. Here we assume that the Euler and continuity equations are valid,\footnote{Since we use the Euler equation here, the following results could equally be obtained with $\Delta z_{\rm Euler}$ in eq.~\eqref{eq:DeltaZ_Euler}, and indeed we find $\langle \overline{\Delta z} \rangle_{R_\perp}^{\rm stack} = \langle \overline{\Delta z}_{\rm Euler} \rangle_{R_\perp}^{\rm stack}$ using the Jeans equation. } leading to the Jeans equation
(see Appendix~\ref{app:Jeans} for a derivation), which in cylindrical coordinates takes the form 
\begin{equation} \label{eq:Jeans}
    \partial_{r_e} \left( \rho \sigma^2_v \right) = - \rho \partial_{r_e} \Psi\, .
\end{equation}

In practice, the individual velocity dispersion of each cluster, $\sigma_v$, cannot be precisely measured due to the limited number of galaxies in each cluster. However, from the width of the stacked redshift distribution, we can extract the line-of-sight velocity dispersion of the stacked clusters as a function of the transverse distance. This quantity, denoted with $\sigma^2_{\rm los} \left( R_\perp \right)$, is related to $\sigma_v$ through
\begin{equation} \label{eq:sigma_measure}
   \sigma_{\rm los}^2 \left( R_\perp \right) \equiv \frac{\int dM \frac{dN}{dM}\int dr_e \rho \sigma^2_v}{\int dM \frac{dN}{dM}\int dr_e \rho } = \frac{\int dM \frac{dN}{dM} \int dr_e r_e \rho \partial_{r_e} \Psi}{{\int dM \frac{dN}{dM}\int dr_e \rho }}\, ,
\end{equation}
where we have used the Jeans equation in the second equality. 
The right-hand side of eq.~\eqref{eq:sigma_measure} is fully determined by $R_s$ and $\rho_0$ (using eq.~\eqref{eq:Poisson_NFW} for $\Psi$) and by the cluster mass function $dN/dM$. Hence, by measuring $\sigma^2_{\rm los}$ at different values of $R_\perp$, one can in principle determine these quantities.

In practice, we proceed slightly differently. 
We first express $\Psi_{\rm NFW}$ from eq.~\eqref{eq:Poisson_NFW} in terms of the concentration parameter $c_v \equiv \frac{R_v}{R_s}$, with $R_v$ the virial radius. We then insert this in the expression for $\sigma_{\rm los}^2 \left( R_\perp \right)$ in eq.~\eqref{eq:sigma_measure}, where we choose as integration variable the virial mass $M_v$, given by
\begin{eqnarray}
\label{eq:virial}
    M_v= 4 \pi \int_0^{R_v} dR R^2 \rho_{\rm NFW} \left( R\right) = \frac{4 \pi R_v^3 \rho_0}{c_v^3} \left( \log \left( 1+ c_v \right) - \frac{c_v}{1+c_v} \right) \, .
\end{eqnarray}
The virial mass can also be related to the overdensity parameter $\delta_c$ through
\begin{eqnarray}
\label{eq:Mvdeltac}
    M_v = \frac{4}{3} \pi R^3_v \delta_c \rho_{\rm crit}\,,
\end{eqnarray}
where $\rho_{\rm crit}=\frac{3H_0^2}{8 \pi G}$ is the critical density of the Universe. Using the measured value of $H_0=67.32 \ {\rm km/s/Mpc}$ from~\cite{Planck:2018vyg} and $\delta_c=102$ as in~\cite{Wojtak:2011ia}, we see that by combining eqs.~\eqref{eq:virial} and~\eqref{eq:Mvdeltac}, we can express $\rho_0$ in terms of $c_v$ and $M_v$ only, and insert it into the NFW potential in eq.~\eqref{eq:Poisson_NFW}. We then assume a simple power-law parametrisation of the mass function as in~\cite{Wojtak:2011ia},\footnote{The slope of the power law $b$ defined in eq.~\eqref{eq:massfunction} is denoted with $\alpha$ in~\cite{Wojtak:2011ia}. We note that there seems to be a typo in the sign of this parameter in~\cite{Wojtak:2011ia}.} 
\begin{equation} \label{eq:massfunction}
    \frac{dN}{dM_v} \propto M_v^{b}\, .
\end{equation}

For a given $R_\perp$, $\sigma_{\rm los}^2$ in eq.~\eqref{eq:sigma_measure} therefore only depends on the concentration parameter $c_v$ and the slope of the mass function $b$. Changing integration variable from $r_e$ to $R$ through $r_e^2=R^2-R_\perp^2$, we obtain
\begin{eqnarray} \label{eq:sigmalosR}
       \sigma_{\rm los}^2 \left( R_\perp \right)
       & =&2  \frac{\int dM_v M_v^{b} \int_{R_\perp}^\infty dR \sqrt{R^2-R_\perp^2} \rho_{\rm NFW} (R) \partial_{R} \Psi_{\rm NFW} (R) }{\int dM_v M_v^{b} \Sigma \left(R_\perp \right)}  \, , \end{eqnarray}
where we have introduced the 2-dimensional surface density profile 
\begin{equation}
\label{eq:Sigma}
   \Sigma (R_\perp)= 2\int_{R_\perp}^\infty dR \frac{R}{\sqrt{R^2- R_\perp^2}}\rho_{\rm NFW} (R) \, .
\end{equation}
By measuring the left-hand side of eq.~\eqref{eq:sigmalosR} at different values of $R_\perp$, we can thus constrain the two free parameters $c_v$ and $b$ entering the right-hand side. In~\cite{Wojtak:2011ia}, the best-fit values are given by $b= -2.3$ and $c_v=5.5$, assuming isotropic galaxy orbits. This allows us to compute numerically the stacked redshift difference in eq.~\eqref{eq:DeltaZ_stack}. Before doing so, however, we model the impact of the BCG position on the signal.

\section{Asymmetry in the BCG position}
\label{sec:BCG_velocity}

So far, we have assumed that the BCG is at rest at the centre of each cluster. In practice, however, the BCG is usually not located exactly at the centre and thus has a non-zero velocity in the cluster reference frame. This impacts the signal in two ways: first, if the BCG is not exactly at the bottom of the cluster gravitational potential, the gravitational redshift difference between the member galaxies and the BCG is reduced; secondly, if the BCG is moving with respect to the centre, the kinematic contribution to the measured signal is modified. This second point has an impact on both the width and the shift of the redshift difference distribution, as the former depends on the linear Doppler contribution and the latter on the second-order Doppler terms in eq.~\eqref{eq:Deltaz_final}. 

\subsection{Impact of the BCG velocity on the width}

We start by computing the impact of the BCG velocity on the width of the redshift distribution. This was previously given by $\sigma^2_{\rm los}$ in eq.~\eqref{eq:sigmalosR}, i.e.~$\big(\overline{\Delta z}(R_\perp)\big)^2$ averaged over the velocity distribution of the member galaxies and integrated over the mass function. This quantity is governed by the linear term $-\Delta v_\parallel=v_{\parallel\,c}-v_{\parallel}(r_e)$ in eq.~\eqref{eq:Deltaz_final}, which takes the form $v_{\parallel\,\rm BCG}-v_{\parallel}(r_e)=-\Delta v_\parallel+\Delta v_{\parallel\,\rm BCG}$ if the BCG is moving. Here, $v_{\rm BCG}$ and $\Delta v_{\rm BCG}=v_{\rm BCG}-v_c$ denote the velocity of the BCG with respect to the observer and the cluster, respectively. The BCG velocity yields a new contribution to the width due to $\left(\Delta v_{\parallel\,\rm BCG}\right)^2$ averaged over the velocity distribution of the BCGs across different clusters. More precisely, by considering that the velocity of the member galaxies is uncorrelated with that of the BCG, the width is given by
\begin{equation}
    \sigma_{\rm tot}^2 = \sigma_{\rm los}^2 + \sigma^2_{\rm BCG}\, , 
\end{equation}
where $\sigma_{\rm BCG}$ is the one-dimensional velocity dispersion of the BCG across the different clusters. 
In eq.~\eqref{eq:sigmalosR}, the width of the distribution is used to infer the parameters $c_v$ and $b$ that determine the profile of the gravitational potential $\Psi_{\rm NFW}$. If the BCG is moving, the observed width is not given by $\sigma_{\rm los}$ but instead by $\sigma_{\rm tot}$, and the left-hand side of eq.~\eqref{eq:sigmalosR} must be replaced with
\begin{equation}
\label{eq:sigmatot}
    \sigma_{\rm los}^2 =   \sigma_{\rm tot}^2 - \sigma^2_{\rm BCG}\, .
\end{equation}

\subsection{Impact of the BCG velocity on the shift}

The BCG velocity also directly affects the shift of the redshift distribution through the second-order Doppler effects in eq.~\eqref{eq:Deltaz_final}. To include the non-vanishing velocity of the BCG in the modelling, we can obtain the redshift difference $\Delta z$ following the steps in section~\ref{subsection:redshift_difference}, but without assuming that the BCG is at rest at the centre of each cluster. In this way, we obtain
\begin{eqnarray}
    \Delta z &= & \HH_e \tilde \Delta r_e - \tilde \Delta \ndv + \frac{\HH_e^2 \left( \tilde \Delta r_e \right)^2 }{2} \left( 1 +\frac{\dot\HH_e}{\HH_e^2} \right) 
+ \HH_e \ndv \tilde \Delta r_e 
+
\tilde \Delta r_e \dot \ndv +\frac{\tilde \Delta v^2}{2} - \tilde \Delta \ndv^2 - \tilde \Delta \Psi  \nonumber \\
&& - \HH_e  \tilde \Delta \ndv \left( 2 \tilde \Delta r_e - r_e\right) + \ndv \tilde \Delta \ndv
- \dot \HH_e r_e \tilde\Delta r_e  
+ \Delta \dot \ndv \left( r_e - \tilde \Delta r_e \right) + \mathcal{O} \left( \epsilon_\HH^3 \right) \, ,
\label{eq:Deltaz_BCG}
\end{eqnarray}
where $\tilde \Delta$ denotes differences between the member galaxies and the BCG.\footnote{When deriving eq.~\eqref{eq:Deltaz_BCG}, we can follow two approaches: we can either expand $\bar{z}_{\rm BCG}$ and $\delta z_{\rm BCG}$ around the emission time of the BCG, which is not assumed to be at the centre, or we can expand all quantities, including $\bar{z}_c$ and $\delta z_c$, around the emission time of a fictitious galaxy located at the centre. Here, we have adopted the second approach. As we will see below, the two approaches lead to the same contribution from $v_{\rm BCG}$. The only contributions that change in the final observable are the background evolution terms proportional to $r_e^2\HH_e^2$, which however do not have an impact on the measured shift of the redshift distribution.} As expected, by assuming that the BCG is at rest at the cluster centre, i.e.~$\tilde \Delta r_e = r_e$ and $\bv_c = \bv_{\rm BCG}$, we recover eq.~\eqref{eq:Deltaz_inter}. 

The galaxy weights $n_g$ in $\overline{\Delta z}(R_\perp)$ from eq.~\eqref{eq:observable_def} are not affected by the BCG velocity, since they only encode the distribution of the member galaxies. Consequently, we can immediately conclude that all terms in eq.~\eqref{eq:Deltaz_BCG} that are linear in the BCG velocity vanish when we average them over the symmetric velocity distribution, following the steps in section \ref{sec:velocity_distribution}. The only contributions that impact the final observable $\langle \overline{\Delta z} \rangle_{R_\perp}^{\rm stack}$ are thus the ones that are quadratic in $v_{\rm BCG}$, namely
\begin{eqnarray}
    \Delta z \supset \frac{\tilde \Delta v^2}{2} -  \tilde \Delta \ndv^2  \supset - \frac{v_{\rm BCG}^2}{2} + {\ndv}_{\rm BCG}^2\,.
\end{eqnarray}
This leads to the following term that should simply be added to $\langle \overline{\Delta z} \rangle_{R_\perp}^{\rm stack}$ in eq.~\eqref{eq:DeltaZ_stack}:
\begin{eqnarray}
    \left\langle  - \frac{v_{\rm BCG}^2}{2} + {\ndv}_{\rm BCG}^2 \right\rangle_{R_\perp}^{\rm stack} = -\frac{1}{2} \sigma^2_{\rm BCG} \, .
\end{eqnarray}
By comparing with eq.~\eqref{eq:DeltaZ_stack} and using the definition of $\sigma^2_{\rm los}$ in eq.~\eqref{eq:sigmalosR}, we remark that $\langle \overline{\Delta z} \rangle_{R_\perp}^{\rm stack}$ has a different dependence on $\sigma^2_{\rm BCG}$ and $\sigma^2_{\rm los}$. First, the contributions from $\sigma_{\rm BCG}$ do not depend on the BCG magnification bias or spectral index. This is expected from the fact that no BCG should be excluded from the catalogue due to fluctuations around the flux threshold of the survey.\footnote{Ref.~\cite{Kaiser:2013ipa} however pointed out that these effects could enhance the flux of another galaxy such that it would become larger than that of the BCG. This would lead to a misidentification of the BCG and therefore bias the measurement. This effect is not included in our theoretical prediction, as it is expected to be small.} Secondly, even the terms that are not proportional to the magnification bias and spectral index are not the same for $\sigma_{\rm los}$ and $\sigma_{\rm BCG}$. This is due to the fact that the observable in eq.~\eqref{eq:def_Deltaz} is not defined symmetrically in $z$ and $z_c$, in order to remove the effect of the expansion of the Universe when stacking clusters.

\subsection{Impact of the BCG position on the shift}

Lastly, if the BCG is not at rest, it will not be located at the bottom of the gravitational potential of the cluster. Consequently, the value of the gravitational potential at the BCG position, $\Psi_{\rm BCG}$,  will differ from the value at the centre of the cluster, $\Psi_c$, thus reducing the amplitude of the observed gravitational redshift signal. 

To compute this effect, we relate the displacement of the BCG from the bottom of the potential to its velocity. Under the assumption of isotropic galaxy orbits, the BCG equation of motion is given by 
\begin{equation}
    \frac{\left( \Delta v_{\rm BCG} \right)^2}{R} =  \partial_R \Psi_{\rm BCG} \, . 
\end{equation}
The NFW potential $\Psi_{\rm NFW} (R)$ in eq.~\eqref{eq:Poisson_NFW} tends to become linear in $R$ close to the centre of the cluster. Hence, we can use the approximation 
\begin{equation}
    \partial_R \Delta \Psi_{\rm BCG} \simeq \frac{\Delta \Psi_{\rm BCG}}{R} \, 
\end{equation}
for the gradient of the gravitational potential difference $\Delta \Psi_{\rm BCG} =\Psi_{\rm BCG} - \Psi_c $, where $\Psi_c=- 4 \pi G R_s^2 \rho_0$. This leads to 
\begin{equation} 
    \left( \Delta v_{\rm BCG} \right)^2 =  R \partial_R \Psi_{\rm BCG} 
    =  R \partial_R \Delta \Psi_{\rm BCG} 
    \simeq  \Delta \Psi_{\rm BCG} \, ,
\end{equation}
such that, by stacking all clusters, we obtain
\begin{equation}
    \langle\left( \Delta v_{\rm BCG} \right)^2 \rangle^{\rm stack} = 3 \sigma^2_{\rm BCG} \simeq \langle   \Delta \Psi_{\rm BCG} \rangle^{\rm stack}\, .
\end{equation}
The gravitational redshift contribution to $\langle\overline{\Delta z}\rangle_{R_\perp}^{\rm stack}$ is then modified as
\begin{eqnarray} \label{eq:DeltaPsi_Stack_BCG}
 \langle   \Delta \Psi \rangle^{\rm stack} &=&
 \langle \Psi \rangle^{\rm stack} -  \langle \Psi_{\rm BCG} \rangle^{\rm stack}= \left(  \langle\Psi \rangle^{\rm stack}-  \langle \Psi_c\rangle^{\rm stack} \right) -  \langle\Delta \Psi_{\rm BCG} \rangle^{\rm stack} 
 \nonumber \\
& \simeq &\left( \langle \Psi \rangle^{\rm stack}- \langle \Psi_c \rangle^{\rm stack}\right) - 3 \sigma^2_{\rm BCG} \, ,
\end{eqnarray}
i.e.~the offset of the BCGs from the bottom of their respective potentials reduces the signal by a term $3 \sigma^2_{\rm BCG}$, in agreement with the result from~\cite{Kaiser:2013ipa}. This effect was neglected in the measurement from~\cite{Wojtak:2011ia}, which only accounted for the impact of the BCG velocity on the width of the distribution through eq.~\eqref{eq:sigmatot}. 

\section{Numerical computation of the shift from stacked clusters}
\label{sec:numerical}
From the expression for $\langle\overline{\Delta z}\rangle_{R_\perp}^{\rm stack}$ in eq.~\eqref{eq:Deltaz_obs}, we remark that the redshift difference between the member galaxies and the BGCs is sourced by three different contributions: one from gravitational redshift, one from the velocity dispersion at second order, and one from the background evolution proportional to $r_e^2\HH_e^2$. We now aim to study the amplitude of these terms and, in particular, to quantify the error induced by interpreting the redshift difference as a pure gravitational redshift effect. In order to compare our theoretical predictions with the measurements of~\cite{Wojtak:2011ia}, we adopt the same setup to model the galaxy clusters: isotropic galaxy orbits, a spherical NFW density profile with concentration parameter $c_v=5.5$, and a cluster mass function described by a power law with exponent $b=-2.3$. These two parameters are needed to predict the amplitude of the signal, and as discussed in section~\ref{sec:model_with_NFW}, they are directly inferred from the measurement of the width of the redshift distribution. They are therefore bound to the catalogue used in~\cite{Wojtak:2011ia}.

\subsection{Gravitational redshift}

The gravitational redshift contribution can be rewritten using $R$ as the integration variable, 
\begin{eqnarray} \label{eq:Deltaz_grav_stack}
    \langle \overline{\Delta z}_{\rm grav} \rangle_{R_\perp}^{\rm stack} 
       & =&- \frac{2\int dM_v M_v^{b} \int_{R_\perp}^\infty dR \frac{R}{\sqrt{R^2- R_\perp^2}}\rho_{\rm NFW} (R)\Delta {\Psi_{\rm NFW} (R)} }{\int dM_v M_v^{-b} \Sigma \left(R_\perp \right)}\nonumber\\
   &&-3 \sigma_{\rm BCG}^2
    {+ \, \mathcal{O} \left( \epsilon_\HH^3 \right)} 
   \, ,
\end{eqnarray}
where $\Sigma$ is given in eq.~\eqref{eq:Sigma}. The quantities $\Delta\Psi_{\rm NFW}$ and $\rho_{\rm NFW}$ are fully determined by $c_v$, which can be derived from measurements of $\sigma^2_{\rm los}$, as discussed in section \ref{sec:model_with_NFW}.

Note that Ref.~\cite{Wojtak:2011ia} inferred $c_v$ and $b$ from $\sigma^2_{\rm tot}$ instead of $\sigma^2_{\rm los}$, leading to an overestimation of the gravitational potential $\Delta\Psi_{\rm NFW}$ in eq.~\eqref{eq:Deltaz_grav_stack}. To correct for this effect, they assumed that the velocity dispersion of the BCG is a fraction of the total one, $\sigma_{\rm BCG}=x\sigma_{\rm tot}$, with fixed $x=0.35$. They then replaced the inferred potential $\Delta\Psi_{\rm NFW}$ in eq.~\eqref{eq:Deltaz_grav_stack} with $(1-x^2)\Delta\Psi_{\rm NFW}$, where the factor follows from eq.~\eqref{eq:sigmatot}. One can then obtain $\rho_{\rm NFW}$ from the Poisson equation in eq.~\eqref{eq:Poisson_NFW} using this same reduced potential. This procedure would be fully correct if $x$ was independent of $R_\perp$. This is however not the case, since $\sigma_{\rm BCG}$  does not depend on the position of the member galaxies and hence it is constant in $R_\perp$, while $\sigma_{\rm tot}$ is not, as can be seen for example in the supplementary figure 2 of~\cite{Wojtak:2011ia}. This implies that $x = \sigma_{\rm BCG}/\sigma_{\rm tot}$ is indeed a function of $R_\perp$. Nevertheless, we have checked that for the range of $R_\perp$ values used in~\cite{Wojtak:2011ia} to infer $c_v$ and $b$ ($R_\perp \lesssim 1 \ {\rm Mpc}$), assuming a constant $x$ is a good approximation. In the following, we will therefore adopt the same constant rescaling $(1-x^2)$, such that we can 
consider the values for $c_v$ and $b$ obtained in~\cite{Wojtak:2011ia} for the sake of comparison.

Lastly, eq.~\eqref{eq:Deltaz_grav_stack} also contains the correction for the BCG offset from the centre $-3\sigma^2_{\rm BCG}$. In this case, assuming a constant value for $x$ is not a good approximation, since the range of $R_\perp$ where the signal is measured is much larger than the one used to infer $c_v$ and $b$. Instead of using a fixed $x$, we therefore consider a fixed $\sigma_{\rm BCG}$ to compute the correction $-3\sigma^2_{\rm BCG}$. We take the value $x=0.35$ at $R_\perp=0.4\, {\rm Mpc}$, which is in the middle of the range considered in the supplementary figure 2 of~\cite{Wojtak:2011ia}. We then take the corresponding value $\sigma_{\rm tot}(R_\perp=0.4\, {\rm Mpc}) \simeq 623 \ {\rm km/s}$, yielding $\sigma_{\rm BCG}\simeq 218\,{\rm km/s}$.

\subsection{Doppler effects}

The Doppler contribution can be directly computed from the measurement of the velocity dispersion $\sigma_{\rm tot}$ and from $\sigma_{\rm BCG}$ inferred above. We have 
\begin{eqnarray} \label{eq:kineticterm}
\langle \overline{\Delta z}_{\rm kin} \rangle_{R_\perp}^{\rm stack} 
&=&         \left(\frac{3}{2} - \mathcal{R} \right) \sigma_{\rm los}^2\left( R_\perp \right) - \frac{1}{2} \sigma^2_{\rm BCG} {+ \, \mathcal{O} \left( \epsilon_\HH^3 \right)} 
\nonumber \\
&=& \left(\frac{3}{2} - \mathcal{R} \right) \left( \sigma_{\rm tot}^2\left( R_\perp \right) - \sigma^2_{\rm BCG}\right)- \frac{1}{2} \sigma^2_{\rm BCG} {+ \, \mathcal{O} \left( \epsilon_\HH^3 \right)} 
    \, .
\end{eqnarray}
$\mathcal{R}$ is given in eq.~\eqref{eq:defR} and depends on the magnification bias $s_b$ and the spectral index $\alpha$. These two quantities can be directly measured from the sample of galaxies, and we will show below the results obtained assuming different values for them (see figure~\ref{fig:grav_vs_kinematics}).

\subsection{Background evolution effects}
\label{subsection:evolution}

\begin{figure}
    \centering
    \includegraphics[width=0.8\linewidth]{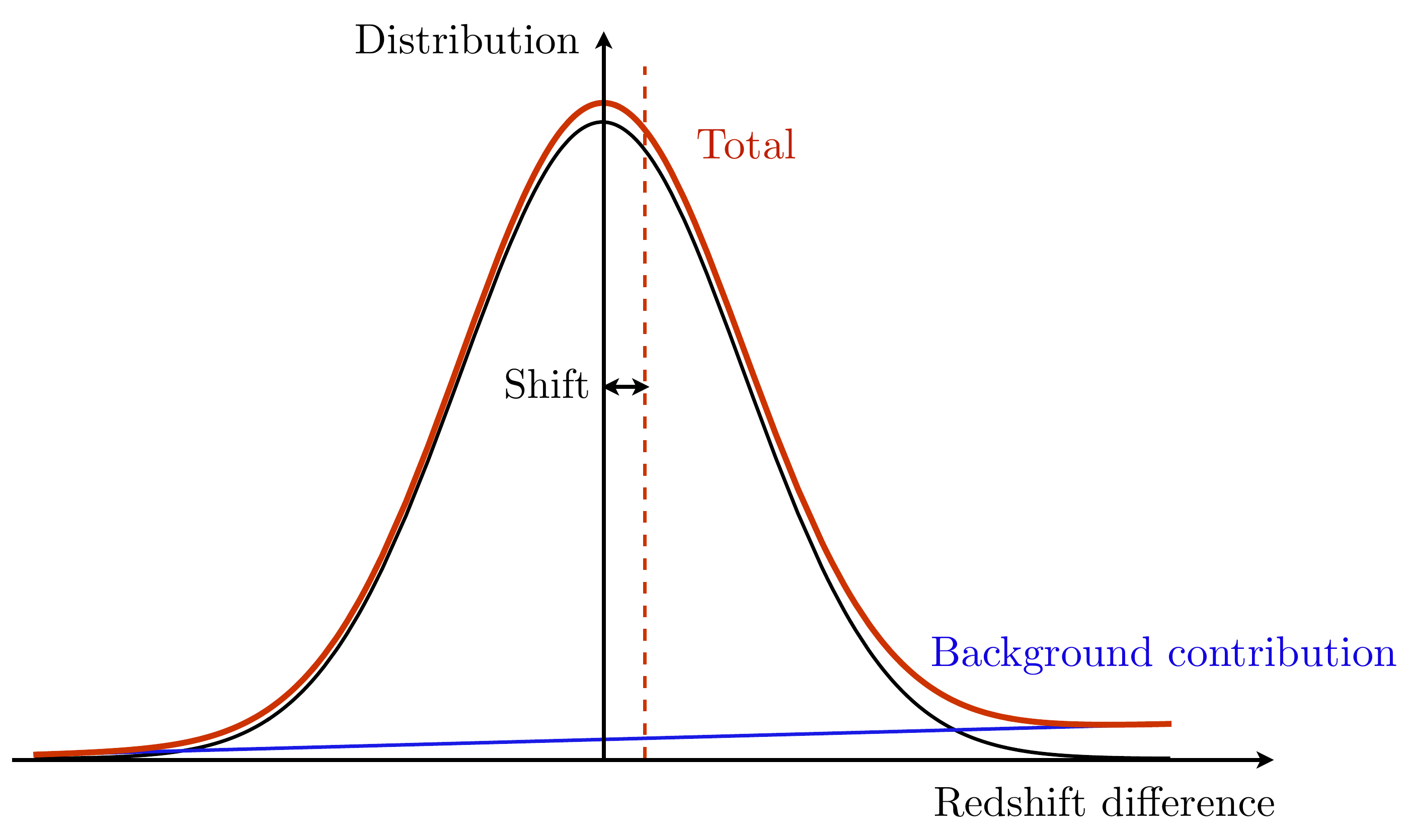}
    \caption{The combination of a Gaussian distribution centred around zero and a linear contribution (generated by the evolution of the background) yields a total contribution with a shifted mean. The effect is exaggerated for illustrative purposes. }
    \label{fig:background_tilt}
\end{figure}

Finally, the last contribution in eq.~\eqref{eq:Deltaz_obs}, proportional to $r_e^2$, is due to the background evolution of the Universe between the emission time of the member galaxy and that of the BCG. This contribution can be removed from the data, such that it does not affect the measured shift of the redshift distribution. As discussed in~\cite{Cai:2016ors}, its impact on the distribution of the redshift difference, i.e.\ on the number of galaxies observed with a given $\Delta z$, is to a good approximation linear in $\Delta z$, as the characteristic timescale of the background evolution is typically set by the Hubble rate. To see this, we consider a perfectly homogeneous universe with density $\bar{\rho}_g\propto 1/a^3$. In this case, the number of galaxy per redshift bin is simply 
\begin{equation}
\label{eq:galaxy_background_Histogram}
        \frac{dN_g}{dz} \propto \frac{\left( 1 + z \right)^2r(z)^2}{\HH} = \frac{(1+z_c)^2 r(z_c)^2}{\HH_e } + \frac{d}{dz} \left( \frac{(1+z)^2 r(z)^2}{\HH } \right)_{z=z_c}\!\!\!\times \Delta z + \mathcal{O}\left( \left(\Delta z\right)^2 \right)\, ,
\end{equation}
where we have performed a Taylor expansion around the emission time of the BCG. For $\Delta z \ll z$, the linear term dominates, such that $dN/dz \propto \Delta z$ (up to an additive constant). As illustrated in figure~\ref{fig:background_tilt}, such a linear contribution to the distribution introduces a shift in the mean of the distribution. Thus, a perfectly Gaussian distribution centred around 0 will be tilted and shifted. The evolution terms that we have computed in eq.~\eqref{eq:Deltaz_obs} are exactly the ones accounting for this shift.

In~\cite{Wojtak:2011ia}, this shift is removed by modelling the distribution as the sum of a Gaussian part with unknown width and shift and a linear function with a free slope. The evolution terms are fully captured by the linear contribution, such that the measured shift is the one of the original distribution, only affected by gravitational redshift and second-order kinematic effects.

\subsection{Numerical results}

\begin{figure}[t!]
\centering
	\includegraphics[width=0.85\columnwidth]{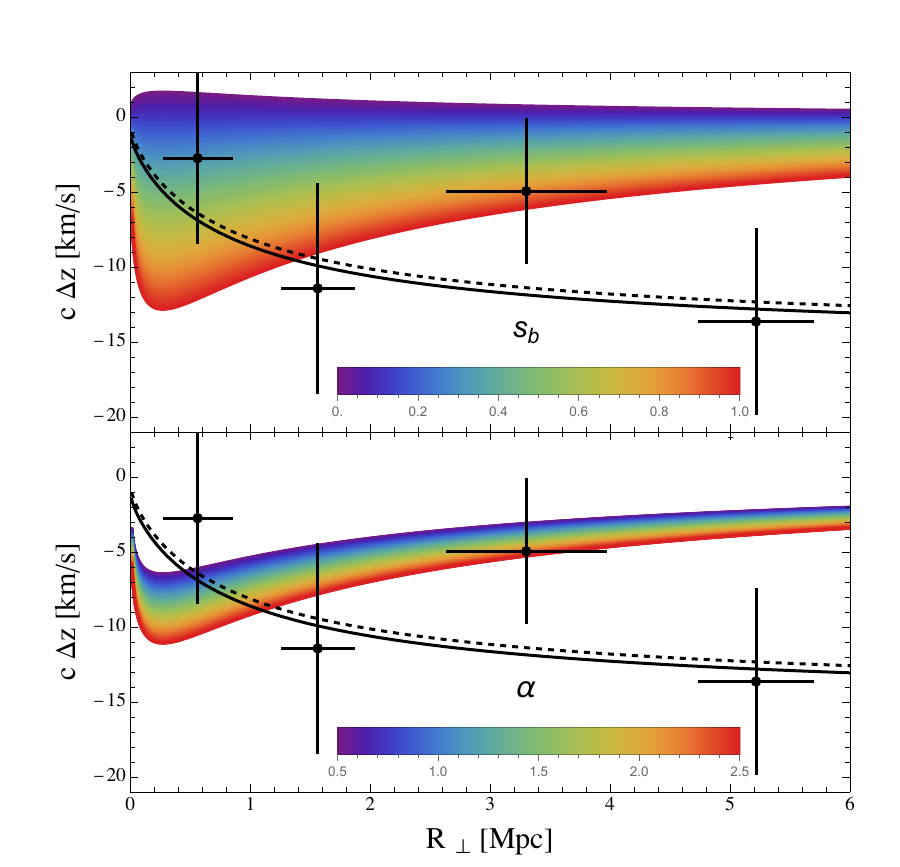}
 \caption{We show the theoretical prediction for gravitational redshift in eq.~\eqref{eq:Deltaz_grav_stack} (black line) and the kinematic effects in eq.~\eqref{eq:kineticterm} (coloured band) as a function of $R_\perp$, together with the measurements from~\cite{Wojtak:2011ia}. The black dashed line corresponds the gravitational redshift contribution without the impact of the BCG motion, i.e.~without the term $-3\sigma^2_{\rm BGC}$ in eq.~\eqref{eq:DeltaPsi_Stack_BCG}. In the top panel, we vary the magnification bias $s_b$ at fixed spectral index $\alpha =2$, while in the bottom panel, we vary $\alpha$ at fixed magnification bias $s_b=0.8$.}
 \label{fig:grav_vs_kinematics} 
 \end{figure}

\begin{figure}[t!]
\centering
\includegraphics[width=0.85\columnwidth]{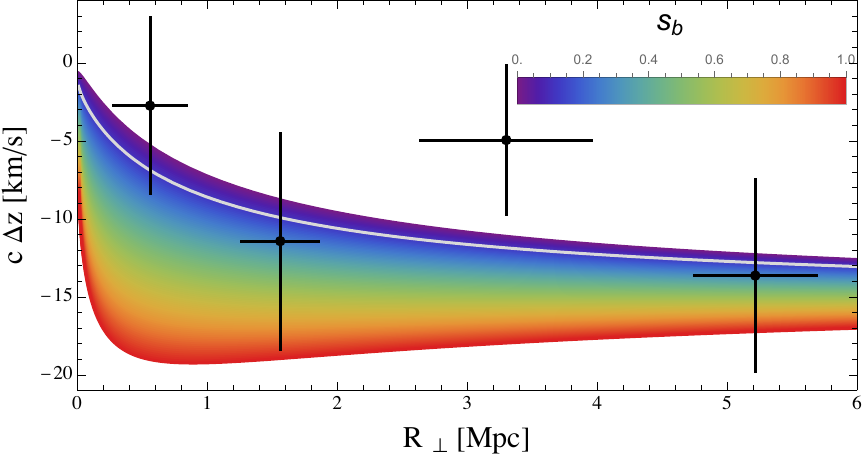}
\caption{We show the theoretical prediction for the total signal, given by the sum of the gravitational redshift and kinematic contributions, for different values of magnification bias from $s_b=0$ (violet) to $s_b=1$ (red), with $   \alpha =2$ as in~\cite{Kaiser:2013ipa}. For comparison, we again plot the measurements from~\cite{Wojtak:2011ia}.
The light gray line indicates the gravitational redshift contribution. 
}
\label{fig:prediction} 
\end{figure}

We can now compare our theoretical prediction with the first detection in clusters from~\cite{Wojtak:2011ia}.
In figure~\ref{fig:grav_vs_kinematics}, we show the four measurements from~\cite{Wojtak:2011ia} with the respective error bars, together with our theoretical predictions as a function of $R_\perp$. We separately plot the gravitational redshift prediction from eq.~\eqref{eq:Deltaz_grav_stack} (black line) and the Doppler contributions given by eq.~\eqref{eq:kineticterm} (colour bands) for various values of the magnification bias $s_b$ and the spectral index $\alpha$. Ref.~\cite{Wojtak:2011ia} only considered gravitational redshift in their theoretical prediction. Their result agrees with ours (in the case of isotropic orbits), except for the contribution $-3\sigma^2_{\rm BCG}$ due to the BCG offset with respect to the centre. Hence, we also show with a dotted black line the prediction without this term, showing that the impact of the BCG offset is relatively small. Comparing the gravitational redshift contribution on its own with the measured signal, we see a good agreement at all scales.

The contamination from the Doppler contributions strongly depends on the value of $s_b$ and $\alpha$. In the top panel of figure~\ref{fig:grav_vs_kinematics}, we show the Doppler contribution for $\alpha=2$, varying the magnification bias $s_b$, while we fix $s_b=0.8$ and vary $\alpha$ in the bottom panel. As already pointed out in previous literature~\cite{Zhao:2012gxk,Kaiser:2013ipa}, these kinematic effects can be comparable to gravitational redshift and therefore cannot be neglected. In an almost volume limited survey, i.e.~where almost all galaxies are detected, the magnification bias is low, and hence the kinematic effects are much smaller than gravitational redshift. However, for more realistic values of the magnification bias, such as $s_b =0.8$ as considered in~\cite{Kaiser:2013ipa}, we find that these effects dominate at small distances from the cluster centre and are still about one third of the gravitational redshift contribution at the largest separation.

Isolating the gravitational redshift contribution relies on an accurate prediction of the kinematic terms, which requires prior knowledge of the magnification bias $s_b$ and a modelling of the spectral index $\alpha$. In building the galaxy catalogue to measure the redshift difference, the individual galaxies must be carefully assigned to the different clusters. Several cuts are applied in this procedure, see for instance~\cite{Sadeh:2014rya}, and we underline that $s_b$ and $\alpha$ must be inferred from the final reduced catalogue.

\begin{figure}[t!]
\centering
\includegraphics[width=0.85\columnwidth]{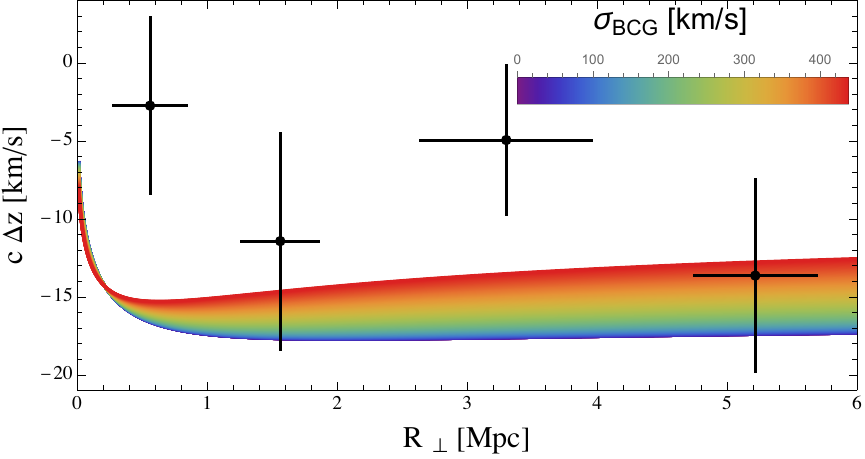} 
\caption{We show the impact of the BCG velocity on the theoretical prediction (colour band), compared with the measurement from~\cite{Wojtak:2011ia}. The BCG velocity impacts both the gravitational redshift contribution through eq.~\eqref{eq:Deltaz_grav_stack} and the kinematic contribution through eq.~\eqref{eq:kineticterm}. The ratio $x=\sigma_{\rm BCG}/\sigma_{\rm tot}$ used in the renormalisation of the gravitational potential $\Psi_{\rm NFW}$ in eq.~\eqref{eq:Deltaz_grav_stack} has been evaluated at $R_\perp= 0.4 \ {\rm Mpc}$, and it varies with $\sigma_{\rm BCG}$.}
\label{fig:sigmaBCG} 
\end{figure}

In figure~\ref{fig:prediction}, we show the total predicted signal for $\alpha=2$ and for different values of $s_b$, together with the measurements from \cite{Wojtak:2011ia}. We see that the measurements are best described by values of $s_b$ close to zero, for which the gravitational redshift contribution strongly dominates over the Doppler terms. Lastly, in figure~\ref{fig:sigmaBCG}, we show how the theoretical prediction varies for different values of the BCG velocity dispersion. As expected from eqs.~\eqref{eq:Deltaz_grav_stack}--\eqref{eq:kineticterm}, a larger BCG velocity dispersion leads to a smaller signal amplitude. In figure \ref{fig:complementary1} in Appendix~\ref{app:figures}, we also show the total prediction for different values of the spectral index $\alpha$ for fixed magnification bias $s_b=0.8$.

\section{Comparison with previous work}
\label{sec:compare}

In this work, we have performed the first derivation of the observed redshift difference $\Delta z$ between the BCG and the other cluster members including all relativistic effects up to second order in the weak-field expansion, given in eq.~\eqref{eq:Deltaz_final}. We have then averaged $\Delta z$ on the light cone by weighting it with the observed galaxy number counts $n_g \left( \bn , z ,F_* \right)$, properly accounting for all the relativistic effects both in $\Delta z$ and $n_g$. We have shown that all linear terms in the weak-field parameter $\epsilon_\HH$ vanish. By the same symmetry argument, we expect the same for all terms proportional to $\epsilon_\HH^3$, such that our result only neglects corrections of order $\epsilon_\HH^4$ and above. 

\begin{figure}[t!]
\centering
\includegraphics[width=0.85\columnwidth]{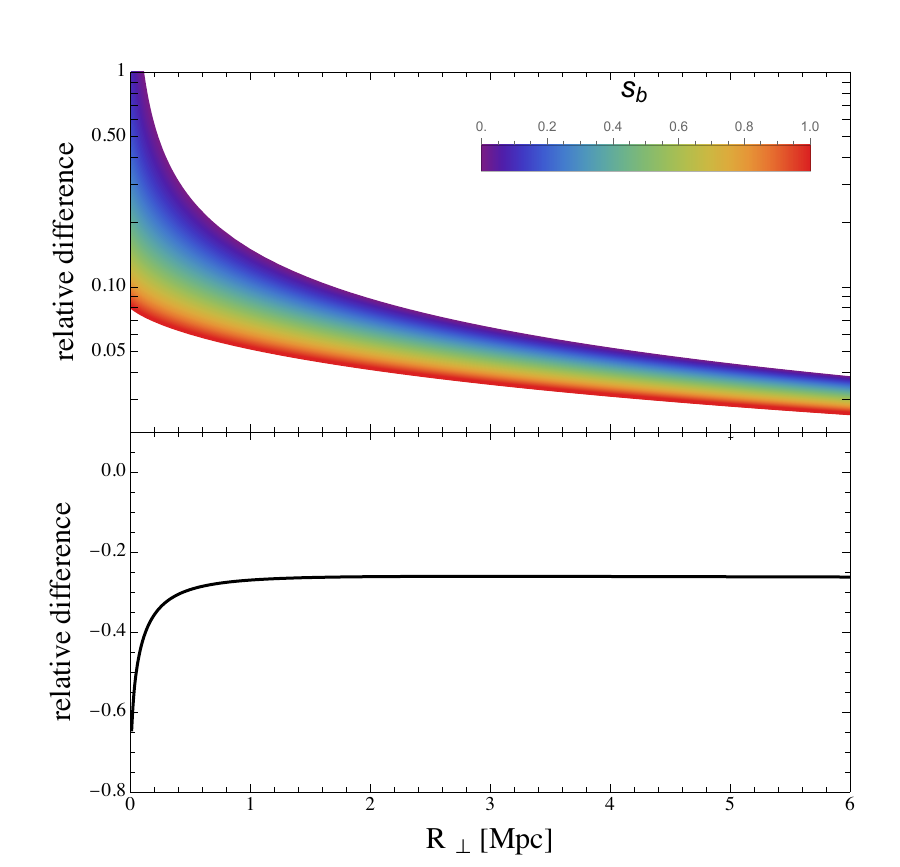}
\caption{We plot the relative difference between our theoretical prediction and K13. In the top panel, we assume that the BCGs are at rest at the bottom of the gravitational potential, i.e.~$\sigma_{\rm BCG} = 0$, and consider different values of the magnification bias parameter $s_b$ and $\alpha=2$. In the bottom panel, we show the relative difference in the contribution from the BCG velocities, setting all the other contributions to zero and fixing the magnification bias $s_b=0.8$ and spectral index $\alpha=2$. We note that the relative difference is independent from the value of $\sigma_{\rm BCG}$.
}
\label{fig:Kaiser_comparison} 
\end{figure}

Various studies have included some of the relativistic corrections, but to the best of our knowledge, none of them properly captures all effects. The transverse Doppler effect was first introduced in~\cite{Zhao:2012gxk}. Ref.~\cite{Kaiser:2013ipa} then added the light-cone and magnification contributions~\cite{Kaiser:2013ipa}. 
All kinematic effects (up to the BCG motion) are summarised in eq.~(8) of~\cite{Kaiser:2013ipa}, denoted with K13 below,
\begin{equation}
\label{eq:Kaiser}
    \langle \overline{\Delta z} \rangle^{(\rm K13)}_{R_\perp} = \left( 2.5 - \left( 3 +    \alpha \right)  \frac{5}{2} s_b \right)\sigma^2_{\rm los } \left( R_\perp \right) = \left( 2.5 - \frac{25}{2} s_b \right) \sigma^2_{\rm los } \left( R_\perp \right)\,,
\end{equation}
where in the last step we have assumed that $\alpha  = 2$, as in~\cite{Kaiser:2013ipa}. Moreover, assuming as in~\cite{Kaiser:2013ipa} that
\begin{eqnarray}
    \frac{d \ln n}{d \ln L}  \simeq -2 \quad\mbox{leading to}\quad s_b \simeq - \frac{2}{5}  \frac{d \ln n}{d \ln L} = \frac{4}{5} = 0.8 \, ,
\end{eqnarray}
the amplitude of the kinematic effects becomes
\begin{eqnarray} \label{eq:Kaiser_short}
       \langle \overline{\Delta z} \rangle^{(\rm K13)}_{R_\perp} = -7.5 \sigma^2_{\rm los } \left( R_\perp \right) \, .
\end{eqnarray}
Our result differs from eq.~\eqref{eq:Kaiser_short} by the absence of the light-cone term. As discussed in section \ref{sec:galaxy_weights}, this is due to the fact that we need to express the density profile on a hypersurface of constant time, as this is the intrinsic frame of the cluster where its density profile can be modelled with the NFW form. As a consequence, the light-cone effect that appears in the galaxy number counts is exactly compensated by the Jacobian arising in the change of coordinate from $\chi$ to $r_e$, $\frac{d \chi}{dr_e}$, applied in eq.~\eqref{eq:Delta_re}. 
Taking all effects into account, our kinematic contribution is given by
\begin{eqnarray}
\langle \overline{\Delta z}_{\rm kin} \rangle_{R_\perp}^{\rm stack}=\left[\frac{3}{2} -\frac{5}{2} s_b \left(3 +   \alpha \right) \right]\sigma^2_{\rm los}\, ,
\end{eqnarray}
which differs from eq.~\eqref{eq:Kaiser} by exactly the light-cone term, $\langle \overline{\Delta z} \rangle^{(\rm LC)}_{R_\perp} = \sigma^2_{\rm los } \left( R_\perp \right)$. 
As we can see from eq.~\eqref{eq:Kaiser_short}, this leads to an overestimation of the amplitude of the kinematic effects of roughly $12\%$ at any separation $R_\perp$, for $s_b=0.8$ and $\alpha = 2$.  
In the top panel of figure~\ref{fig:Kaiser_comparison}, we show the relative difference between our prediction and K13 (for $\sigma_{\rm BCG}=0$), for fixed $\alpha=2$, and varying $s_b$. We see that at small separation $R_\perp\sim 0.5$\,Mpc, the difference is non-negligible, ranging between 6 and 25\% depending on the value of $s_b$. When the separation increases, the relative difference becomes less and less relevant. This is simply due to the fact that at large separations the signal is dominated by gravitational redshift, so that the difference between our Doppler contamination and that in K13 does not play an important role.
In figure~\ref{fig:complementary2} in Appendix~\ref{app:figures}, we show the relative difference for different values of the spectral index $\alpha$ for fixed magnification bias $s_b=0.8$. 

Another difference with respect to K13 lies in our treatment of the BCG velocity and its impact on the kinematic contribution. As discussed in section~\ref{sec:BCG_velocity}, there is an asymmetry between the BCG velocity and that of the members in the definition of the observable in eq.~\eqref{eq:observable_def}, which is due to the fact that the redshift difference needs to be rescaled by the BCG redshift when stacking clusters at different redshifts. This leads to an asymmetry between the velocity dispersions of the BCG and the members in the final stacked signal. K13 instead includes a correction for the BCG velocity with the same numerical pre-factor as that in front of the velocity dispersion of the members in eq.~\eqref{eq:Kaiser}, i.e.~subtracting $-5/2\sigma^2_{\rm BCG}$ from the kinematic contribution. Moreover, K13 assumes a constant ratio between $\sigma_{\rm BCG}$ and $\sigma_{\rm los}$ implying that $\sigma_{\rm BCG}$ depends on $R_\perp$. As discussed in section~\ref{sec:numerical}, this does not impact the determination of $c_v$ and $b$, which is performed across a sufficiently small interval in $R_\perp$. However, the assumption of a constant ratio leads to an erroneous non-negligible variation of $\sigma_{\rm BCG}$ over the large range of $R_\perp$ where the signal is measured. In our work, we instead consider a constant $\sigma_{\rm BCG}$. In the bottom panel of figure~\ref{fig:Kaiser_comparison}, we show the relative difference between our treatment of the BCG motion and that of K13. To perform this comparison, we choose $\sigma_{\rm BCG}=218 \ {\rm km/s}$ at $R_\perp=0.4 \ {\rm Mpc}$ in both cases and compute the BCG prediction from K13, using the corresponding value $x=0.35$. We see that the two predictions differ by 30 to 60\% for the $s_b=0.8$ and $\alpha=2$. 

We can also compare our results with~\cite{Cai:2016ors}, where the redshift difference $\Delta z$ is derived up to first order in the metric perturbation $\Psi$ and second order in the peculiar velocity $v^2$, assuming the validity of the Euler equation and without performing the density weighting. Our expression for $\Delta z$ in eq.~\eqref{eq:Deltaz_final} instead does not rely on the Euler equation and can therefore be used to test its validity. Using the Euler equation to compare with the result in~\cite{Cai:2016ors}, denoted with $\Delta z^{\rm (C16)}$, we found that the latter missed a quadratic term in the velocity coupled to the background expansion, given by $-r_e \bv \cdot \nabla \ndv$. Hence, we have
\begin{equation}
    \Delta z^{(\rm C16)} = \Delta z_{\rm Euler} + r_e \bv \cdot \nabla \ndv 
    \, ,
\end{equation}
where $\Delta z_{\rm Euler}$ denotes our result in the case where the Euler equation is valid, given in eq.~\eqref{eq:DeltaZ_Euler}. The missing term cannot be neglected, as it is of the same order as the linear ones in the weak-field expansion, since $\nabla v \sim \delta \sim \mathcal{O} \left( \epsilon^0 \right)$ is of order unity.
Using eq.~\eqref{eq:Euler}, our expression relates to C16 by
\begin{align}
    \Delta z^{(\rm C16)} &= \Delta z_{\rm Euler} -    r_e \dot \ndv -r_e \HH \ndv + r_e \partial_{r_e} \Psi\\
    &=\Delta z_{\rm Euler} -    r_e \Delta\dot{v_\parallel} -r_e \HH\Delta \ndv - r_e\dot {\ndv}_c - r_e\HH {\ndv}_c + r_e \partial_{r_e} \Psi\, .\nonumber
\end{align}
All the additional terms are of second order in the weak-field expansion, therefore contributing to the final observable through
\begin{eqnarray}
    \langle \overline{\Delta z}^{(\rm C16)} \rangle_{R_\perp}^{\rm stack} &\supset& \int d^3\left( \Delta v \right) f \left( \boldsymbol{\Delta }\bv \right) \int dr_e r_e \rho_g^{\rm real } \left( \eta_e , r_e \right)
    \left( \partial_{r_e} \Psi - \Delta \dot \ndv - \HH \Delta \ndv - \dot {\ndv}_c - \HH {\ndv}_c \right)
        \nonumber \\
    &=&  \int d^3\left( \Delta v \right) f \left( \boldsymbol{\Delta }\bv \right) \int dr_e r_e \rho_g^{\rm real } \left( \eta_e , r_e \right)
    \partial_{r_e} \Psi \, ,
\end{eqnarray}
where we have used the fact that the terms linear in $\ndv$ and $v_{\parallel c}$ vanish once integrated over the velocity distribution. We can then relate the derivative of the gravitational potential $\Psi$ to the velocity dispersion through the Jeans equation in eq.~\eqref{eq:Jeans}. Using eq.~\eqref{eq:sigma_measure}, we thus find 
\begin{equation}
    \langle \overline{\Delta z}^{(\rm C16)} \rangle_{R_\perp}^{\rm  stack} = \langle \overline{\Delta z}_{\rm Euler} \rangle_{R_\perp}^{\rm stack} + \sigma^2_{\rm los} \left( R_\perp \right) \, ,
\end{equation}
implying that ref.~\cite{Cai:2016ors} overestimates the kinematic contribution by roughly $ 12 \%$.

\section{Conclusion}
\label{sec:conclusion}

In this paper, we have developed a robust and model-independent modelling of the redshift difference between the BGC and the member galaxies in clusters. The mean value of this difference, weighted over the galaxy distribution, is of particular interest in cosmology, since it is directly sensitive to gravitational redshift, but is not affected by the linear Doppler contribution. This provides a robust method to measure the former effect, which is otherwise strongly subdominant. However, as pointed out in previous literature \cite{Zhao:2012gxk,Kaiser:2013ipa}, second-order Doppler effects also contribute to the mean, contaminating the measurement of gravitational redshift. Various studies have derived some of these second-order contributions. In our work, we have used the relativistic framework for large-scale structure developed in~\cite{Yoo:2009au, Yoo:2010ni, Challinor:2011bk, Bonvin:2011bg,Jeong:2011as,Yoo:2014sfa, Bertacca:2014dra, DiDio:2014lka,DiDio:2016ykq,DiDio:2020jvo} to consistently derive all contributions at second order in the weak-field expansion. This expansion is valid for any value of the density contrast and respects the correct hierarchy between the gravitational potential and the velocities in virialised objects like clusters. Within this framework, we have derived expressions for both the redshift difference $\Delta z$ and the galaxy distribution $n_g$, which are fully model-independent i.e.~do not rely on the Euler equation nor on a specific theory of gravity.

We have found that the contamination from Doppler effects is given by the sum of two types of contributions. The first one arises from second-order effects in $\Delta z$, multiplying the zeroth-order galaxy distribution $n_g$. The second contribution contains the product of the linear $\Delta z$ and the linear galaxy distribution $n_g$. This relies on the relativistic expression for $n_g$ at linear order in the weak-field expansion, which accounts for all projection effects. However, we have shown that most of the terms in $n_g$ cancel out in the final result, due to the integration over the cluster members. More precisely, all displacement terms like RSD and the light-cone effect vanish and only the magnification bias remains. This differs from the result in~\cite{Kaiser:2013ipa}, where RSD do not contribute to the final expression but the light-cone effect still appears.

Comparing the amplitude of gravitational redshift with that of second-order Doppler terms, we remark that the importance of the latter strongly depends on the properties of the galaxy sample, through the magnification bias parameter $s_b$ and the spectral index $\alpha$. These parameters thus need to be precisely determined for the specific galaxy sample used in the measurement in order to model the Doppler contamination. In addition, it is also necessary to predict the velocity dispersion of the BCG, which reduces both the gravitational redshift contribution and the Doppler one when the BCG is not exactly at rest at the centre of the cluster. We have found that the resulting signal reduction is smaller than predicted in the original work of~\cite{Wojtak:2011ia}. 
Our correction due to the BCG motion is instead larger than in~\cite{Kaiser:2013ipa} and has a different scaling in $R_\perp$. 

Since the amplitude of gravitational redshift depends on the theory of gravity, our expression for the shift can in principle be used to test gravity at the scale of clusters. However, this requires particular care. While our expressions for $\Delta z$ and $n_g$ are fully general and valid in any metric theory of gravity, we have assumed the validity of general relativity to predict the amplitude of the gravitational redshift contribution. This assumption enters because the signal depends on the density profile of the cluster. Here, we have considered a NFW profile, which depends on the concentration parameter $c_v$. To determine this parameter from the data, one needs to assume the Jeans equation to relate the velocity dispersion in the cluster to the gravitational potential $\Psi$ and the Poisson equation to relate $\Psi$ to the density, also assuming that the two potentials $\Phi$ and $\Psi$ are the equal. With this procedure, $c_v$ can be inferred from the measurement of the width of the redshift distribution. In the presence of deviations from general relativity, the Poisson equation and the link between $\Phi$ and $\Psi$ are typically modified, and this procedure would yield a wrong measurement of $c_v$. Hence, to properly test the theory of gravity, it is necessary to perform these steps allowing for deviations from general relativity. 

Lastly, since our derivation of $\langle\overline{\Delta z}\rangle^{\rm stack}_\perp$ in eq.~\eqref{eq:DeltaZ_stack} does not rely on the validity of the Euler equation, it can be used to test this equation by comparing the inferred gravitational potential with galaxy velocities, similarly to what has been proposed in the linear regime~\cite{Bonvin:2018ckp,Castello:2024lhl}. This is of particular interest since galaxies are mostly constituted of dark matter, for which the validity of the Euler equation remains a crucial open question. Typically, in models where dark matter interacts non-gravitationally with another component in the Universe, such as dark energy or dark radiation, the Euler equation is violated. In a forthcoming paper, we will present a method to perform this test on cluster scales using the theoretical prediction for gravitational redshift derived in this work.


\section*{Acknowledgments}
We thank Radoslaw Wojtak for sharing their redshift difference measurements from~\cite{Wojtak:2011ia}. We acknowledge support from the European Research Council (ERC) under the European Union's Horizon 2020 research and innovation program (grant agreement No.~863929; project title ``Testing the law of gravity with novel large-scale structure observables'').
 \newpage
\appendix
\section{Galaxy number counts up to first order in the weak-field expansion}
\label{app:weakfield_numbercounts}

The galaxy number counts as a function of the line-of-sight direction $\bn$ and the observed redshift $z$ are defined as 
\begin{eqnarray}
    \Delta \left( \bn , z \right) = \frac{n_g \left( \bn , z \right) - \langle n_g \rangle \left( z \right) }{\langle n_g \rangle \left( z \right)} = \left( 1+\delta_z \left( \bn , z \right)  \right) \frac{\mathcal{V} \left( \bn , z \right)}{\mathcal{\bar V} \left( z \right)} -1 \, .
\end{eqnarray}
Here, $\langle .. \rangle$ denotes the angular average at fixed observed redshift $z$, $n_g\left( \bn, z \right) = dN/dz/d\Omega$ is the number density of sources and $\mathcal{V} \left( \bn , z \right)$ is the volume density per redshift and solid angle. The density perturbation in redshift space is given by 
\begin{eqnarray}
    \delta_z \left( \bn , z \right) &=& \frac{\rho_g \left( \bn , z \right) - \langle \rho_g \rangle \left( z \right)}{\langle \rho_g \rangle \left( z \right)}  = \frac{\bar \rho_g \left( \bar z \right) \left(1 + \delta\left( \bar z \right) \right) - \bar \rho_g \left( z \right)}{ \bar \rho_g\left( z \right)} 
    \nonumber \\
    &=&
    \left( 1 + \delta \right) \left( 1 - \frac{d \log \bar \rho_g}{d \bar z} \delta z \right) -1 + \mathcal{O} \left( \epsilon_\HH^2 \right) \, , 
\end{eqnarray}
where all terms are evaluated at the background redshift $\bar z$.
The volume perturbation to first order in the weak-field approximation has already been computed in~\cite{DiDio:2020jvo}:
\begin{eqnarray}
    \frac{\mathcal{V} \left( \bn ,z \right) }{\mathcal{\bar V} \left( z \right)} = \left( 1- \frac{d \log \mathcal{\bar V}}{d \bar z} \delta z \right) \left( 1+\frac{d \delta z }{ d z} \right)^{-1} \left( 1 + \frac{\delta z}{1+ \bar z } \right)
    {  + \mathcal{O} \left( \epsilon_\HH^2 \right) }
    \, ,
\end{eqnarray}
leading to
\begin{equation}
    \Delta \left( \bn , z \right) = \left( 1 + \delta \right) \left( 1 - \frac{d \log \bar \rho_g}{d \bar z} \delta z \right) \left( 1- \frac{d \log \mathcal{\bar V}}{d \bar z} \delta z \right) \left( 1+\frac{d \delta z }{ d z} \right)^{-1} \left( 1 + \frac{\delta z}{1+ \bar z } \right)
    {  + \mathcal{O} \left( \epsilon_\HH^2 \right) }\, .
\end{equation}
All contributions to this expression have a clear physical meaning: the first factor contains the density contrast, while the second and the third one encode the change in the background number density and the background volume due to the shift in redshift $\delta z$; the fourth factor contains the Jacobian term arising from changing from the background to the observed redshift, leading to RSD, while the last factor is the so-called light-cone term. We can now use
\begin{eqnarray} \label{eq:dlogrhodz}
    \frac{d}{d \bar z} \log  \bar \rho  &=& \frac{3-f_{\rm evo}}{1+\bar z} \, , 
\\ 
\label{eq:dlogVdz}
    \frac{ d \log \mathcal{\bar V}}{d \bar z} &=&\left( 1+ \bar z \right)^{-1} \left( \frac{\dot\HH}{\HH^2} + \frac{2}{\HH \chi} -4 \right) \, , 
\end{eqnarray}
where we have introduced the evolution bias $f_{\rm evo}$ that will be further discussed in section~\ref{app:magnification_bias}. This yields 
\begin{eqnarray}
\label{eq:A7}
        \Delta (\bn, z ) &=&
      \left( 1+ \delta \right)  \left( 1 + \frac{d \delta z}{d\bar z} \right)^{-1}
     \left[1 +  \left( 2 + f_{\rm evo} -\frac{\dot\HH}{\HH^2} - \frac{2}{\HH \chi}  \right)  \frac{\delta z }{ 1 + \bar z  } \right]
 -1 
 + \mathcal{O} \left( \epsilon_\HH^2 \right)  
 \, . \qquad
\end{eqnarray}

\subsection{Magnification bias}
\label{app:magnification_bias}
In the derivation above, we have assumed that we can observe all the galaxies at a given redshift $z$. In practice, however, real surveys are flux limited, meaning that galaxies can be detected only if their observed flux is above a given threshold $F_*$. Hence, the observed number counts are also a function of $F_*$. The observed flux limit $F_*$ and the intrinsic luminosity threshold $L_* (z)$ at the observed redshift $z$ are related by $F_*= \frac{L_*({ \bn },z)}{4 \pi D_L ({ \bn }, z)^2}$, where $D_L$ denotes the luminosity distance. Since $D_L$ depends on direction (due to inhomogeneities~\cite{Bonvin:2005ps,Hui:2005nm}), the intrinsic luminosity threshold corresponding to a fixed flux threshold also depends on direction. We can thus perform a Taylor expansion of the galaxy number counts at fixed observed flux around the background luminosity threshold  $\bar L_*$, yielding
\begin{eqnarray}
    n_g \left( \bn ,z, F_* \right) &=& n_g   \left( \bn ,z, L_* \left( \bn, z \right)  \right) =
n_g \left( \bn , z , \bar L_* \left( z \right) \right)\left( 1+ \left. \frac{\partial \ln  n_g}{\partial \ln L} \right|_{L = \bar L_*} \frac{\delta L_*}{\bar L_*} \right) + \mathcal{O} \left( \epsilon_\HH^2 \right)
    \nonumber  \\
    &=& \bar n_g \left( z, \bar L_* \left(z \right) \right) \left( 1 +  \Delta \left(\bn , z,  \bar L_* \left( z \right) \right)\right) \left( 1 + \left. \frac{\partial \ln  n_g}{\partial \ln L} \right|_{L = \bar L_*} \frac{\delta L_*}{\bar L_*} 
    \right) + \mathcal{O} \left( \epsilon_\HH^2 \right) \, . \qquad
\end{eqnarray}
From this, we find 
\begin{eqnarray}
    \Delta \left(\bn , z,  L_* \left(\bn,  z \right) \right) &=& \frac{n_g \left( \bn, z,  L_* \left(z \right) \right) - \bar n_g \left( z, \bar L_* \left(z \right) \right) }{\bar n_g \left( z, \bar L_* \left(z \right) \right)} 
    \nonumber \\
    &=&  \left( 1 +  \Delta \left(\bn , z,  \bar L_* \left( z \right) \right)\right) \left( 1 + \left. \frac{\partial \ln  n_g}{\partial \ln L} \right|_{L = \bar L_*} \frac{\delta L_*}{\bar L_*} 
    \right)   + \mathcal{O} \left( \epsilon_\HH^2 \right) \, .
\end{eqnarray}
Considering that $\delta L_*$ is linear in the weak-field parameter (it is proportional to the luminosity distance fluctuations, as discussed below), we only need to consider the 0th-order expression for the galaxy number counts appearing in the logarithmic derivative, yielding\footnote{In cases where a galaxy bias term is included, its dependence on luminosity has to be carefully considered. Similarly, if the density contrast $\delta$ is treated non-perturbatively, one needs to account for its luminosity dependence as well.}
\begin{eqnarray}
    \Delta \left( \bn , z , L_*(\bn,z)\right) 
   & =&\left( 1 +  \Delta \left(\bn , z,  \bar L_* \left( z \right) \right)\right) \left( 1 + \left. \frac{\partial \ln  \bar n_g}{\partial \ln L} \right|_{L = \bar L_*} \frac{\delta L_*}{\bar L_*} 
    \right)   + \mathcal{O} \left( \epsilon_\HH^2 \right)\\
    &=&
    \Delta \left( \bn , z , \bar L_* (z)\right) + \left( 1 + \delta \right) \left( 1+ \frac{d \delta z}{d \bar z} \right)^{-1}\left. \frac{\partial \ln  \bar n_g}{\partial \ln L} \right|_{L= \bar{L}_*} \frac{\delta L_*}{\bar{L}_*} +\mathcal{O} \left( \epsilon_\HH^2 \right)
    \, . \qquad \nonumber
\end{eqnarray}
This flux cut of the galaxy catalogue introduces a selection bias, denominated magnification bias and defined as 
\begin{eqnarray}
\label{eq:def_sb}
    s_b = \left. - \frac{2}{5} \frac{\partial \ln \bar n_g}{\partial \ln L} \right|_{L= \bar {L}_*} \, .
\end{eqnarray}
By considering that the luminosity fluctuation is twice the luminosity distance fluctuation at fixed flux, i.e.~$\frac{\delta L_*}{\bar{L}_*} = 2 \frac{\delta D_L}{\bar{D}_L}$, we obtain
\begin{eqnarray} \label{eq:Delta_with_sb}
    \Delta \left( \bn , z , L_*(\bn,z)\right)  &=& \Delta \left( \bn , z, \bar L_* (z)\right) 
        -5 s_b   \left( 1 + \delta \right) \left( 1+ \frac{d \delta z}{d \bar z} \right)^{-1} \frac{\delta D_L}{\bar{D}_L}  +\mathcal{O} \left( \epsilon_\HH^2 \right) \, .
\end{eqnarray}
Therefore, to include the magnification bias term, we need to compute the luminosity distance fluctuation up to first order in the weak-field expansion. We start by considering the reciprocity relation between the luminosity and the angular diameter distance~\cite{Etherington},
\begin{eqnarray}
    D_L = D_A \left( 1 + z \right)^2 \, ,
\end{eqnarray}
which leads to 
\begin{eqnarray} 
\label{eq:deltaDL_deltaDA}
      \frac{\delta D_L}{\bar{D}_L} =   \frac{\delta D_A}{\bar{D}_A} \, .
\end{eqnarray}
The angular distance is determined from the infinitesimal area element in the source frame as~\cite{Jeong:2011as,Yoo:2016vne}
\begin{eqnarray}
        dA= D_A^2 d\Omega_o= \sqrt{-g} \epsilon_{\mu \nu \alpha \beta} u^\mu N^\nu \frac{\partial x^\alpha}{\partial \theta} \frac{\partial x^\beta}{\partial \varphi} d\theta d\varphi \, ,
\end{eqnarray}
with
\begin{eqnarray}
    N^\mu = \frac{k^\mu}{u_\nu k^\nu} + u^\mu.
\end{eqnarray}
Here, $k^{\mu}$ is the photon 4-momentum and $N^{\mu}$ is its projection on the hypersurface of constant time for the source, i.e.~$N^\mu u_\mu=0$. To first order in the weak-field approximation, the 4-velocity is given by\footnote{The light-like vector $k^\mu$ solves the geodesic equation. In the conformal metric, the Christoffel symbols are at order $\epsilon_\HH$ or higher. The Christoffel symbols at order $\epsilon_\HH$ are spatial derivatives of the metric perturbations, so by writing them as total and partial time derivatives, and integrating over the affine parameter, we obtain $\delta k^\mu = \mathcal{O}\left( \epsilon_\HH^2 \right)$.}
\begin{eqnarray}
    \left( u ^\mu \right) = a^{-1} \left( 1 , \bv \right) \qquad \text{and} \qquad     \left( k^\mu \right) = a^{-2}\left( 1 , \bn \right) \, , 
\end{eqnarray}
yielding
\begin{eqnarray}
    \left(N^\mu \right) =a^{-1} \left(- \ndv, 1,v_\theta, v_\varphi \right) +\mathcal{O} \left( \epsilon_\HH^2 \right)
\end{eqnarray}
and 
\begin{eqnarray}
        D_A^2 = \frac{dA}{d\Omega}   &=&  a^4 \chi^2  \left( u^0 N^\chi -u^\chi N^0 \right)= \left[  a^2 \chi^2  \right]_{\bar z}  
        = \left[  a^2 \chi^2  \right]_{ z} \left( 1 - \frac{d \log a^2 \chi^2}{dz} \delta z\right) 
    \nonumber \\
    &=&
    \left[  a^2 \chi^2  \right]_{ z} \left( 1 -\left( \frac{2}{\HH \chi} -2 \right)\frac{\delta z}{1+ z}\right)  \, .
\end{eqnarray}
This leads to
\begin{equation} \label{eq:distance_fluctuation}
    \frac{\delta D_L}{\bar D_L} = \frac{\delta D_A}{\bar D_A} 
    =\left( 1 - \frac{1}{\HH \chi} \right) \frac{\delta z}{1+z} +\, \mathcal{O} \left( \epsilon_\HH^2 \right) 
    \, .
\end{equation}
We can thus insert this in eq.~\eqref{eq:Delta_with_sb}, yielding the expression for the galaxy number counts including magnification bias:
\begin{eqnarray} \label{eq:numbercounts_with_magnbias}
\Delta \left( \bn , z , L_*(\bn,z)\right)  &=& \Delta \left( \bn , z \right) 
        -5 s_b  \left( 1 + \delta \right) \left( 1+ \frac{d \delta z}{d \bar z} \right)^{-1} \left( 1 - \frac{1}{\HH \chi} \right) \frac{\delta z}{1+z}  + \mathcal{O} \left( \epsilon_\HH^2 \right)  
\nonumber\\
&=&        
     \left( 1+ \delta \right) \left( 1 + \frac{d \delta z}{d\bar z} \right)^{-1}
    \!\! \left[1\! + \! \left(2 + f_{\rm evo} - \frac{\dot\HH}{\HH^2} -5s_b + \frac{5s_b - 2}{\HH \chi}  \right)   \frac{\delta z }{ 1 + \bar z  }\right]\!\! -1 
     \nonumber \\
&&
+\, \mathcal{O} \left( \epsilon_\HH^2 \right)     
        \, ,
\end{eqnarray}
where all the perturbations are evaluated at the background redshift $\bar z$.

Eq.~\eqref{eq:numbercounts_with_magnbias} depends on the evolution bias $f_{\rm evo}$, which in a flux-limited survey is given by 
\begin{eqnarray}
     \frac{\partial \log  \bar \rho_g}{\partial \bar z}   &=& \frac{3-f_{\rm evo}}{1+\bar z} \, ,
\end{eqnarray}
where the total derivative of eq.~\eqref{eq:dlogrhodz} is replaced by a partial derivative due to the dependence on the luminosity (which depends on redshift itself).
From here, we notice that the evolution bias depends on the evolution of the underlying galaxy population, regardless of the properties of the observing instrument, as discussed in~\cite{Challinor:2011bk}. However, if the number of observed galaxies depends on the luminosity threshold, we can relate the evolution bias $f_{\rm evo}$ to a total derivative by including a term accounting for the change in the luminosity threshold with respect to the redshift, see e.g.~refs.~\cite{Nadolny:2021hti,Maartens:2021dqy}:
\begin{eqnarray}
    f_{\rm evo} &=&  3- \left( 1+ \bar z \right)   \frac{\partial \log  \bar \rho_g}{\partial \bar z} =  3- \left( 1+ \bar z \right)   \frac{d \log  \bar \rho_g}{d \bar z} + \left( 1+ \bar z \right) \frac{\partial \log \bar \rho_g}{\partial \bar L_*} \frac{d \bar L_*}{d \bar z} 
    \nonumber \\
    &=& 3- \left( 1+ \bar z \right)   \frac{d \log  \bar \rho_g}{d \bar z} - \frac{5}{2} s_b  \left( 1 +    \alpha + \frac{2}{\HH \chi} \right) \, . \label{eq:f_evo_alpha}
\end{eqnarray}
Here, we have assumed a power-law frequency dependent luminosity function $L \propto \nu^{-   \alpha}$ and used~\cite{Dalang:2021ruy} 
\begin{eqnarray} \label{eq:def_alpha}
    \frac{d \bar L_*}{d \bar z} = \left( 1 +     \alpha + \frac{2}{\HH \chi} \right) \frac{\bar L_*}{1+ \bar z} \, .
\end{eqnarray}
In this expression, we have chosen to compute the derivative of the luminosity threshold $\bar{L}_*$ with respect to the background redshift $\bar{z}$ at fixed flux density at the source, since our aim is to express $\rho^{\rm real}_g$ in $\overline{\Delta z}$ in terms of intrinsic quantities in the source frame that can be modelled through local physics. This differs from the usual derivations of $\Delta$, where the aim is instead to express it in terms of observable quantities, i.e.\ in terms of the observed flux density, see e.g.~\cite{Maartens:2021dqy,Bonvin:2023jjq}.

We can then insert eq.~\eqref{eq:f_evo_alpha} into eq.~\eqref{eq:numbercounts_with_magnbias}, obtaining
\begin{eqnarray}
\label{eq:numbercounts_final}
  &&  \hspace{-0.7cm}\Delta \left( \bn , z, F_*\right) 
    \nonumber \\
&=& \left( 1+ \delta \right) \left( 1 + \frac{d \delta z}{d\bar z} \right)^{-1}
     \left[1 - \left( \frac{d\log \bar \rho_g}{d \bar z}+ \frac{d\log \mathcal{\bar V}}{d \bar z} \right) \delta z + \left( 1  - \frac{5}{2} s_b \left(3 +   \alpha\right)  \right) \frac{\delta z}{1+\bar z}\right] -1 
     \nonumber \\
&&
+ \mathcal{O} \left( \epsilon_\HH^2 \right) 
\nonumber \\
&=& \left( 1+ \delta \right) \left( 1 + \frac{d \delta z}{d\bar z} \right)^{-1} \left( 1- \frac{d \log \bar n_g}{d \bar z} \delta z\right)
     \left[1 + \left( 1 - \frac{5}{2} s_b \left(3 +   \alpha \right) \right) \frac{\delta z}{1+\bar z}\right] -1 
+ \mathcal{O} \left( \epsilon_\HH^2 \right) 
\, ,
\nonumber \\
\end{eqnarray} 
where we have used eq.~\eqref{eq:dlogVdz}. This is our final result for the galaxy number counts up to first order in the weak-field expansion for flux-limited surveys. We remark that, while this expression describes the observed number counts as a function of the observed redshift $z$, all perturbations on the right-hand side are evaluated at the background redshift $\bar{z}$, allowing for a straightforward integration along the line of sight in eq.~\eqref{eq:2.18}.

\section{Galaxy number counts in real space}
\label{app:Delta_realspace}

After having derived the galaxy number counts up to first order in the weak-field expansion, we perform a change of coordinates from the observed redshift $z$ to the background comoving distance~$\chi = \chi\left( \bar z \right) $. We emphasise that the resulting number counts $\Delta (\boldsymbol{\chi})$ are not observable and thus are not required to be gauge-invariant or free from infrared divergences, unlike the observed counts, $\Delta (z, \bn)$ (see~\cite{Jeong:2011as,Scaccabarozzi:2018vux,Grimm:2020ays,Castorina:2021xzs}).
To perform the change of variable, we start from the fact that the number of sources is preserved in this mapping,\footnote{We remark that $n_g \left(  z, \bn, F_* \right)$ and $n_g \left( \chi, \bn,L_* \right)$ have different units: the first is dimensionless, while the latter has units of inverse length.} 
\begin{equation} \label{eq:change_delta}
    dz \ n_g \left( \bn, z ,F_* \right) = d\chi \ n_g \left(\mathbf{n},\chi,F_* \right) \, .
\end{equation}
With this, we obtain
\begin{eqnarray}
\label{eq:Delta_rcoord}
    \Delta \left( \boldsymbol{\chi},  L_* \right) &=& \frac{dz}{d \chi} \frac{\bar n_g \left( z, \bar L_* \right) }{\bar n_g \left( \chi ,\bar L_*\right)}  \left( 1+ \Delta \left( \bn, z  ,L_*\right) \right) -1 
    { + \mathcal{O}\left( \epsilon_\HH^2 \right)}
    \nonumber \\
    &=&
    \frac{d\bar z}{d\chi} \frac{\bar n_g \left( \bar z,\bar L_* \right)}{\bar n_g \left( \chi ,\bar L_*\right)} 
    \left( 1 + \frac{d \delta z}{d \bar z} \right) \left( 1+ \frac{d \log \bar n_g}{d \bar z} \delta z \right)\left( 1 + \Delta \left(\bn, z ,L_*\right) \right) -1 
        { + \mathcal{O}\left( \epsilon_\HH^2 \right)}
    \nonumber \\ 
    &=& 
 \left(  1+ \delta \right)       \left[1 + \left( 1 - \frac{5}{2} s_b \left(3 +   \alpha \right) \right) \frac{\delta z}{1+\bar z}\right]
 -1
     { + \mathcal{O}\left( \epsilon_\HH^2 \right)}
       \nonumber \\ 
    &=& 
 \left(  1+ \delta \right)       \left[1 + \left(  \frac{5}{2} s_b \left(3 +   \alpha \right) -1 \right) \ndv\right]-1 
     { + \mathcal{O}\left( \epsilon_\HH^2 \right)}\, ,
\end{eqnarray}
where $\boldsymbol{\chi} = -\chi \bn $ and  we have used the background solution of eq.~\eqref{eq:change_delta}
\begin{equation} \label{eq:B3}
    \bar n_g \left( \chi ,\bar L_*\right) d\chi = \bar n_g \left( \bar z,\bar L_* \right) d \bar z \, .
\end{equation}
The velocity term in eq.~\eqref{eq:Delta_rcoord} is the well-known light-cone effect~\cite{Kaiser:2013ipa,Bonvin:2013ogt}, induced by not describing the galaxy density fluctuation in a 3-dimensional box but along the past light-cone. The other contribution in eq.~\eqref{eq:Delta_rcoord} is a selection effect modulated by the source peculiar velocity. The number counts in real space appear in the observable when integrating along the line of sight $\chi$, see eq.~\eqref{eq:2.18}.

\section{Jeans equation}
\label{app:Jeans}
In this section, we highlight the relation between the Jeans equation and the continuity and Euler equations, which describe the motion of individual galaxies.
By modelling the galaxy distribution as an ideal pressure-less fluid with stress-energy tensor $T^{\mu \nu} = \rho u^\mu u^\nu$, we derive, from the  covariant conservation of the energy-momentum tensor $\nabla_\mu T^{\mu \nu}=0$, the continuity ($\nu=0$) and the Euler ($\nu=i$) equations in GR. In the weak-field approximation,
the continuity and the Euler equations take the form
\begin{eqnarray}
    \dot \rho + 3 \HH \rho + \nabla \cdot \left( \rho \bv \right) = \mathcal{O}\left(\epsilon_\HH^2\right) \, , \\
    \dot \bv + \left(\bv \cdot \nabla \right) \bv + \HH \bv + \nabla \Psi = \mathcal{O}\left(\epsilon_\HH^3\right)  \, ,
\end{eqnarray}
where the left-hand sides contain the term present in the Newtonian approximation, while the right-hand sides incorporate the effects of the GR dynamics.
We first multiply the Euler equation by $\rho$, 
\begin{eqnarray}
    && \rho \dot \bv + \rho \left(\bv \cdot \nabla \right) \bv + \rho \HH \bv + \rho  \nabla \Psi = \mathcal{O} \left( \epsilon_\HH^3 \right) 
        \nonumber \\
    &\Rightarrow& \partial_\eta \left(\rho \bv \right) - \dot \rho \bv   + \rho \left(\bv \cdot \nabla \right) \bv + \rho \HH \bv + \rho  \nabla \Psi = \mathcal{O} \left( \epsilon_\HH^3 \right)    
            \nonumber \\
    &\Rightarrow& \partial_\eta \left(\rho \bv \right)    + \rho \left(\bv \cdot \nabla \right) \bv + 4 \rho \HH \bv + \bv \left(\nabla \cdot \left( \rho \bv \right) \right)+ \rho  \nabla \Psi = \mathcal{O} \left( \epsilon_\HH^3 \right) , 
\end{eqnarray}
where in the last step we have replaced $\dot \rho$ using the continuity equation. This can also be written as
\begin{eqnarray} \label{eq:C4}
 &&   \partial_\eta \left(\rho v^j \right) + \rho v^i \partial_i v^j + 4 \rho \HH v^j + \partial_i \left(\rho v^i \right) v^j + \rho \partial_j \Psi =\mathcal{O} \left( \epsilon_\HH^3 \right)
 \nonumber \\
 &\Rightarrow&
    \partial_\eta \left(\rho v^j \right) + \partial_i \left(\rho v^i  v^j \right) + 4 \rho \HH v^j + \rho \partial_j \Psi =\mathcal{O} \left( \epsilon_\HH^3 \right) \, .
\end{eqnarray}
Multiplying the continuity equation with the velocity $v^j$ and the velocity distribution in the cluster reference frame, $f(v)$, we then obtain
\begin{eqnarray} \label{eq:C5}
 &&   \dot \rho \langle v^j \rangle + 3 \HH \rho \langle v^j \rangle + \langle v^j \rangle \partial_i \left(\rho \langle v^i \rangle \right)=\mathcal{O} \left( \epsilon_\HH^3 \right) 
 \nonumber \\
 &\Rightarrow&
  \dot \rho \langle v^j \rangle + 3 \HH \rho \langle v^j \rangle +  \partial_i \left(\rho \langle v^i \rangle \langle v^j \rangle \right) - \rho \langle v^i \rangle \partial_i \langle v^j \rangle=\mathcal{O} \left( \epsilon_\HH^3 \right) .
\end{eqnarray}
We now can repeat the same steps with eq.~\eqref{eq:C4},
\begin{eqnarray}
 &&   \dot \rho \langle v^j \rangle +    \rho \langle \dot v^j \rangle + \partial_i \left(\rho \langle v^i v^j \rangle \right) + 4 \rho \HH \langle v^j \rangle+ \rho \partial_j \Psi =\mathcal{O} \left( \epsilon_\HH^3 \right)
    \nonumber \\
   &\Rightarrow&
        \rho \langle \dot v^j \rangle + \partial_i \left(\rho \langle v^i v^j \rangle - \rho \langle v^i\rangle \langle v^j \rangle \right) +  \rho \HH \langle v^j \rangle+ \rho \partial_j \Psi +  \rho \langle v^i \rangle \partial_i \langle v^j \rangle 
        \nonumber \\
        &&
        + \left[\dot \rho \langle v^j \rangle  +  \partial_i \left(\rho \langle v^i\rangle \langle v^j \rangle \right) + 3 \rho \HH \langle v^j \rangle   - \rho \langle v^i \rangle \partial_i \langle v^j \rangle \right]=\mathcal{O} \left( \epsilon_\HH^3 \right)
         \nonumber \\
   &\Rightarrow&
        \rho \langle \dot v^j \rangle + \partial_i \left(\rho \sigma^2_{ij}\right) +  \rho \HH \langle v^j \rangle+ \rho \partial_j \Psi +  \rho \langle v^i \rangle \partial_i \langle v^j \rangle =\mathcal{O} \left( \epsilon_\HH^3 \right)  \, ,
\end{eqnarray}
where the terms in the squared brackets vanishes due to eq.~\eqref{eq:C5}.

We are interested in expressing the Jeans equation in cylindrical coordinates $\left\{R_\perp,Z,\varphi \right\} $ for a static system, i.e.~neglecting the time derivatives and taking $\langle v_\perp \rangle = \langle v_Z \rangle =0$.  Moreover, in a virialised system at equilibrium, we can assume that the velocities along different axes are uncorrelated and that the velocity dispersion along the radial and vertical directions are the same, $\langle v_\perp^2 \rangle =\langle v_Z^2 \rangle = \sigma^2$, such that the stress tensor is diagonal 
\begin{eqnarray}
    \left( \sigma_{ij} \right)^2= \left( \begin{array}{ccc} 
\sigma^2_v & 0 & 0\\
0& \sigma^2_v & 0 \\
0 & 0 & \langle  v_\varphi^2 \rangle -\langle  v_\varphi \rangle^2 
\end{array}
    \right) \, .
\end{eqnarray}
Under these assumptions, the Jeans equation along the $Z$-direction simplifies to 
\begin{eqnarray} \label{eq:jeansZ}
   \partial_Z \left( \rho \sigma^2_v \right) + \rho \partial_Z \Psi =\mathcal{O} \left( \epsilon_\HH^3 \right)  \, .
\end{eqnarray}
It is worth remarking that, when working at second order in the weak-field expansion, the dynamics is fully captured by the Newtonian approximation.
In the main text in eq.~\eqref{eq:Jeans}, the derivative along $Z$ is replaced by the derivative along $r_e$, using $Z=\chi_{\rm BCG}+r_e$.

\section{Additional figures}
\label{app:figures}
In this section, we reproduce some figures presented in the main text, varying the spectral index $\alpha$ instead of the magnification bias $s_b$. 

\begin{figure}[t!]
\centering
	\includegraphics[width=0.85\columnwidth]{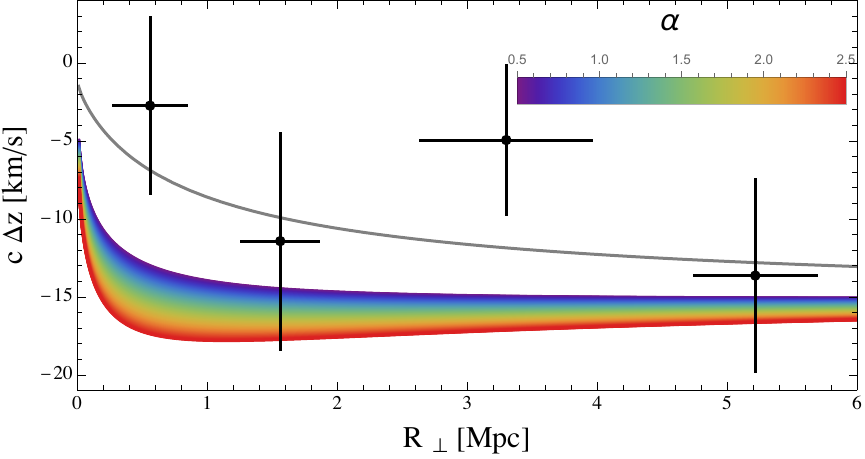}
 \caption{Figure analogous to the top panel of figure~\ref{fig:prediction}, but varying the spectral index $\alpha$ from $0.5$ (violet) to $2.5$ (red), while keeping the magnification bias fixed to $s_b = 0.8$. }
 \label{fig:complementary1} 
 \end{figure}
 \begin{figure}[t!]
\centering
	\includegraphics[width=0.85\columnwidth]{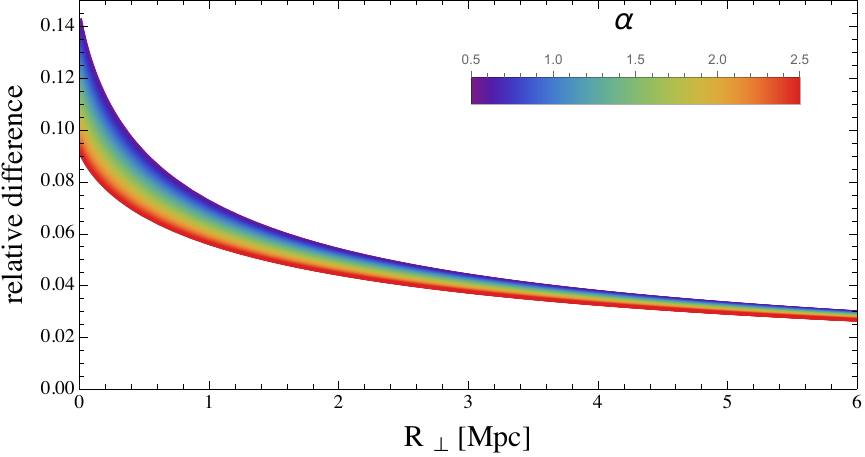}
 \caption{Figure analogous to figure~\ref{fig:Kaiser_comparison}, but varying the spectral index $\alpha$ from $0.5$ (violet) to $2.5$ (red), while keeping the magnification bias fixed to $s_b = 0.8$.}
 \label{fig:complementary2} 
 \end{figure}


\bibliography{grav_red_biblio}

\providecommand{\href}[2]{#2}\begingroup\raggedright\begin{thebibliography}{10}

\bibitem{Einstein1907}
A.~Einstein, {\it {\"U}ber das relativit{\"a}tsprinzip und die aus demselben
  gezogenen folgerungen},  {\em Jahrbuch der Radioaktivit{\"a}t und Elektronik}
  {\bf 4} (1907) 411--462.

\bibitem{Pound:1959PhR}
R.~V. {Pound} and G.~A. {Rebka}, {\it {Gravitational Red-Shift in Nuclear
  Resonance}},  {\em PRL.} {\bf 3} (Nov., 1959) 439--441.

\bibitem{Lopresto:1991oxy}
J.~C. {Lopresto}, C.~{Schrader}, and A.~K. {Pierce}, {\it {Solar Gravitational
  Redshift from the Infrared Oxygen Triplet}},  {\em APJ.} {\bf 376} (Aug.,
  1991) 757.

\bibitem{Alam:2017izi}
S.~Alam, H.~Zhu, R.~A.~C. Croft, S.~Ho, E.~Giusarma, and D.~P. Schneider, {\it
  {Relativistic distortions in the large-scale clustering of SDSS-III BOSS
  CMASS galaxies}},  {\em Mon. Not. Roy. Astron. Soc.} {\bf 470} (2017), no.~3
  2822--2833, [\href{http://arxiv.org/abs/1709.07855}{{\tt arXiv:1709.07855}}].

\bibitem{Nottale:1990}
L.~Nottale, {\it Gravitational redshifts and lensing by large scale
  structures},  in {\em Gravitational Lensing} (Y.~Mellier, B.~Fort, and
  G.~Soucail, eds.), (Berlin, Heidelberg), pp.~29--38, Springer Berlin
  Heidelberg, 1990.

\bibitem{Cappi:1995}
A.~{Cappi}, {\it {Gravitational redshift in galaxy clusters.}},  {\em Astronomy
  and Astrophysics} {\bf 301} (Sept., 1995) 6.

\bibitem{Kim:2004tc}
Y.-R. Kim and R.~A.~C. Croft, {\it {Gravitational redshifts in simulated galaxy
  clusters}},  {\em Astrophys. J.} {\bf 607} (2004) 164--174,
  [\href{http://arxiv.org/abs/astro-ph/0402047}{{\tt astro-ph/0402047}}].

\bibitem{Wojtak:2011ia}
R.~Wojtak, S.~H. Hansen, and J.~Hjorth, {\it {Gravitational redshift of
  galaxies in clusters as predicted by general relativity}},  {\em Nature} {\bf
  477} (2011) 567--569, [\href{http://arxiv.org/abs/1109.6571}{{\tt
  arXiv:1109.6571}}].

\bibitem{Sadeh:2014rya}
I.~Sadeh, L.~L. Feng, and O.~Lahav, {\it {Gravitational Redshift of Galaxies in
  Clusters from the Sloan Digital Sky Survey and the Baryon Oscillation
  Spectroscopic Survey}},  {\em Phys. Rev. Lett.} {\bf 114} (2015), no.~7
  071103, [\href{http://arxiv.org/abs/1410.5262}{{\tt arXiv:1410.5262}}].

\bibitem{Jimeno:2014xma}
P.~Jimeno, T.~Broadhurst, J.~Coupon, K.~Umetsu, and R.~Lazkoz, {\it {Comparing
  gravitational redshifts of SDSS galaxy clusters with the magnified redshift
  enhancement of background BOSS galaxies}},  {\em Mon. Not. Roy. Astron. Soc.}
  {\bf 448} (2015), no.~3 1999--2012,
  [\href{http://arxiv.org/abs/1410.6050}{{\tt arXiv:1410.6050}}].

\bibitem{eBOSS:2021ofn}
{\bf eBOSS} Collaboration, C.~T. Mpetha et~al., {\it {Gravitational redshifting
  of galaxies in the SPIDERS cluster catalogue}},  {\em Mon. Not. Roy. Astron.
  Soc.} {\bf 503} (2021), no.~1 669--678,
  [\href{http://arxiv.org/abs/2102.11156}{{\tt arXiv:2102.11156}}].

\bibitem{Rosselli:2022qoz}
D.~Rosselli, F.~Marulli, A.~Veropalumbo, A.~Cimatti, and L.~Moscardini, {\it
  {Testing general relativity: New measurements of gravitational redshift in
  galaxy clusters}},  {\em Astron. Astrophys.} {\bf 669} (2023) A29,
  [\href{http://arxiv.org/abs/2206.05313}{{\tt arXiv:2206.05313}}].

\bibitem{Zhao:2012gxk}
H.~Zhao, J.~A. Peacock, and B.~Li, {\it {Testing gravity theories via
  transverse Doppler and gravitational redshifts in galaxy clusters}},  {\em
  Phys. Rev. D} {\bf 88} (2013), no.~4 043013,
  [\href{http://arxiv.org/abs/1206.5032}{{\tt arXiv:1206.5032}}].

\bibitem{Kaiser:2013ipa}
N.~Kaiser, {\it {Measuring Gravitational Redshifts in Galaxy Clusters}},  {\em
  Mon. Not. Roy. Astron. Soc.} {\bf 435} (2013) 1278,
  [\href{http://arxiv.org/abs/1303.3663}{{\tt arXiv:1303.3663}}].

\bibitem{Cai:2016ors}
Y.-C. Cai, N.~Kaiser, S.~Cole, and C.~Frenk, {\it {Gravitational redshift and
  asymmetric redshift-space distortions for stacked clusters}},  {\em Mon. Not.
  Roy. Astron. Soc.} {\bf 468} (2017), no.~2 1981--1993,
  [\href{http://arxiv.org/abs/1609.04864}{{\tt arXiv:1609.04864}}].

\bibitem{Yoo:2009au}
J.~Yoo, A.~L. Fitzpatrick, and M.~Zaldarriaga, {\it {A New Perspective on
  Galaxy Clustering as a Cosmological Probe: General Relativistic Effects}},
  {\em Phys. Rev. D} {\bf 80} (2009) 083514,
  [\href{http://arxiv.org/abs/0907.0707}{{\tt arXiv:0907.0707}}].

\bibitem{Yoo:2010ni}
J.~Yoo, {\it {General Relativistic Description of the Observed Galaxy Power
  Spectrum: Do We Understand What We Measure?}},  {\em Phys. Rev. D} {\bf 82}
  (2010) 083508, [\href{http://arxiv.org/abs/1009.3021}{{\tt
  arXiv:1009.3021}}].

\bibitem{Challinor:2011bk}
A.~Challinor and A.~Lewis, {\it {The linear power spectrum of observed source
  number counts}},  {\em Phys. Rev. D} {\bf 84} (2011) 043516,
  [\href{http://arxiv.org/abs/1105.5292}{{\tt arXiv:1105.5292}}].

\bibitem{Bonvin:2011bg}
C.~Bonvin and R.~Durrer, {\it {What galaxy surveys really measure}},  {\em
  Phys. Rev. D} {\bf 84} (2011) 063505,
  [\href{http://arxiv.org/abs/1105.5280}{{\tt arXiv:1105.5280}}].

\bibitem{Jeong:2011as}
D.~Jeong, F.~Schmidt, and C.~M. Hirata, {\it {Large-scale clustering of
  galaxies in general relativity}},  {\em Phys. Rev. D} {\bf 85} (2012) 023504,
  [\href{http://arxiv.org/abs/1107.5427}{{\tt arXiv:1107.5427}}].

\bibitem{Yoo:2014sfa}
J.~Yoo and M.~Zaldarriaga, {\it {Beyond the Linear-Order Relativistic Effect in
  Galaxy Clustering: Second-Order Gauge-Invariant Formalism}},  {\em Phys. Rev.
  D} {\bf 90} (2014), no.~2 023513, [\href{http://arxiv.org/abs/1406.4140}{{\tt
  arXiv:1406.4140}}].

\bibitem{Bertacca:2014dra}
D.~Bertacca, R.~Maartens, and C.~Clarkson, {\it {Observed galaxy number counts
  on the lightcone up to second order: I. Main result}},  {\em JCAP} {\bf 09}
  (2014) 037, [\href{http://arxiv.org/abs/1405.4403}{{\tt arXiv:1405.4403}}].

\bibitem{DiDio:2014lka}
E.~Di~Dio, R.~Durrer, G.~Marozzi, and F.~Montanari, {\it {Galaxy number counts
  to second order and their bispectrum}},  {\em JCAP} {\bf 12} (2014) 017,
  [\href{http://arxiv.org/abs/1407.0376}{{\tt arXiv:1407.0376}}]. [Erratum:
  JCAP 06, E01 (2015)].

\bibitem{DiDio:2016ykq}
E.~Di~Dio, F.~Montanari, A.~Raccanelli, R.~Durrer, M.~Kamionkowski, and
  J.~Lesgourgues, {\it {Curvature constraints from Large Scale Structure}},
  {\em JCAP} {\bf 06} (2016) 013, [\href{http://arxiv.org/abs/1603.09073}{{\tt
  arXiv:1603.09073}}].

\bibitem{DiDio:2020jvo}
E.~Di~Dio and F.~Beutler, {\it {The relativistic galaxy number counts in the
  weak field approximation}},  {\em JCAP} {\bf 09} (2020) 058,
  [\href{http://arxiv.org/abs/2004.07916}{{\tt arXiv:2004.07916}}].

\bibitem{Bonvin:2018ckp}
C.~Bonvin and P.~Fleury, {\it {Testing the equivalence principle on
  cosmological scales}},  {\em JCAP} {\bf 05} (2018) 061,
  [\href{http://arxiv.org/abs/1803.02771}{{\tt arXiv:1803.02771}}].

\bibitem{Bonvin:2020cxp}
C.~Bonvin, F.~O. Franco, and P.~Fleury, {\it {A null test of the equivalence
  principle using relativistic effects in galaxy surveys}},  {\em JCAP} {\bf
  08} (2020) 004, [\href{http://arxiv.org/abs/2004.06457}{{\tt
  arXiv:2004.06457}}].

\bibitem{Umeh:2020cag}
O.~Umeh, K.~Koyama, and R.~Crittenden, {\it {Testing the equivalence principle
  on cosmological scales using the odd multipoles of galaxy cross-power
  spectrum and bispectrum}},  {\em JCAP} {\bf 08} (2021) 049,
  [\href{http://arxiv.org/abs/2011.05876}{{\tt arXiv:2011.05876}}].

\bibitem{Bonvin:2022tii}
C.~Bonvin and L.~Pogosian, {\it {Modified Einstein versus Modified Euler for
  Dark Matter}},  {\em Nature Astron.} {\bf 7} (2023), no.~9 1127--1134,
  [\href{http://arxiv.org/abs/2209.03614}{{\tt arXiv:2209.03614}}].

\bibitem{Castello:2024jmq}
S.~Castello, Z.~Wang, L.~Dam, C.~Bonvin, and L.~Pogosian, {\it {Disentangling
  modified gravity from a dark force with gravitational redshift}},  {\em Phys.
  Rev. D} {\bf 110} (2024), no.~10 103523,
  [\href{http://arxiv.org/abs/2404.09379}{{\tt arXiv:2404.09379}}].

\bibitem{Castello:2024lhl}
S.~Castello, Z.~Zheng, C.~Bonvin, and L.~Amendola, {\it {Testing the
  equivalence principle across the Universe: A model-independent approach with
  galaxy multitracing}},  {\em Phys. Rev. D} {\bf 111} (2025), no.~12 123559,
  [\href{http://arxiv.org/abs/2412.08627}{{\tt arXiv:2412.08627}}].

\bibitem{Kehagias:2013rpa}
A.~Kehagias, J.~Nore\~na, H.~Perrier, and A.~Riotto, {\it {Consequences of
  Symmetries and Consistency Relations in the Large-Scale Structure of the
  Universe for Non-local bias and Modified Gravity}},  {\em Nucl. Phys. B} {\bf
  883} (2014) 83--106, [\href{http://arxiv.org/abs/1311.0786}{{\tt
  arXiv:1311.0786}}].

\bibitem{Creminelli:2013nua}
P.~Creminelli, J.~Gleyzes, L.~Hui, M.~Simonovi\'c, and F.~Vernizzi, {\it
  {Single-Field Consistency Relations of Large Scale Structure. Part III: Test
  of the Equivalence Principle}},  {\em JCAP} {\bf 06} (2014) 009,
  [\href{http://arxiv.org/abs/1312.6074}{{\tt arXiv:1312.6074}}].

\bibitem{Kesden:2006zb}
M.~Kesden and M.~Kamionkowski, {\it {Galilean Equivalence for Galactic Dark
  Matter}},  {\em Phys. Rev. Lett.} {\bf 97} (2006) 131303,
  [\href{http://arxiv.org/abs/astro-ph/0606566}{{\tt astro-ph/0606566}}].

\bibitem{Desmond:2020gzn}
H.~Desmond and P.~G. Ferreira, {\it {Galaxy morphology rules out
  astrophysically relevant Hu-Sawicki $f(R)$ gravity}},  {\em Phys. Rev. D}
  {\bf 102} (2020), no.~10 104060, [\href{http://arxiv.org/abs/2009.08743}{{\tt
  arXiv:2009.08743}}].

\bibitem{Breton:2018wzk}
M.-A. Breton, Y.~Rasera, A.~Taruya, O.~Lacombe, and S.~Saga, {\it {Imprints of
  relativistic effects on the asymmetry of the halo cross-correlation function:
  from linear to non-linear scales}},  {\em Mon. Not. Roy. Astron. Soc.} {\bf
  483} (2019), no.~2 2671--2696, [\href{http://arxiv.org/abs/1803.04294}{{\tt
  arXiv:1803.04294}}].

\bibitem{Bonvin:2005ps}
C.~Bonvin, R.~Durrer, and M.~A. Gasparini, {\it {Fluctuations of the luminosity
  distance}},  {\em Phys. Rev. D} {\bf 73} (2006) 023523,
  [\href{http://arxiv.org/abs/astro-ph/0511183}{{\tt astro-ph/0511183}}].
  [Erratum: Phys.Rev.D 85, 029901 (2012)].

\bibitem{Hui:2005nm}
L.~Hui and P.~B. Greene, {\it {Correlated Fluctuations in Luminosity Distance
  and the (Surprising) Importance of Peculiar Motion in Supernova Surveys}},
  {\em Phys. Rev. D} {\bf 73} (2006) 123526,
  [\href{http://arxiv.org/abs/astro-ph/0512159}{{\tt astro-ph/0512159}}].

\bibitem{Navarro:1995iw}
J.~F. Navarro, C.~S. Frenk, and S.~D.~M. White, {\it {The Structure of cold
  dark matter halos}},  {\em Astrophys. J.} {\bf 462} (1996) 563--575,
  [\href{http://arxiv.org/abs/astro-ph/9508025}{{\tt astro-ph/9508025}}].

\bibitem{Bonvin:2013ogt}
C.~Bonvin, L.~Hui, and E.~Gaztanaga, {\it {Asymmetric galaxy correlation
  functions}},  {\em Phys. Rev. D} {\bf 89} (2014), no.~8 083535,
  [\href{http://arxiv.org/abs/1309.1321}{{\tt arXiv:1309.1321}}].

\bibitem{Bonvin:2014owa}
C.~Bonvin, {\it {Isolating relativistic effects in large-scale structure}},
  {\em Class. Quant. Grav.} {\bf 31} (2014), no.~23 234002,
  [\href{http://arxiv.org/abs/1409.2224}{{\tt arXiv:1409.2224}}].

\bibitem{Bonvin:2015kuc}
C.~Bonvin, L.~Hui, and E.~Gaztanaga, {\it {Optimising the measurement of
  relativistic distortions in large-scale structure}},  {\em JCAP} {\bf 08}
  (2016) 021, [\href{http://arxiv.org/abs/1512.03566}{{\tt arXiv:1512.03566}}].

\bibitem{Gaztanaga:2015jrs}
E.~Gaztanaga, C.~Bonvin, and L.~Hui, {\it {Measurement of the dipole in the
  cross-correlation function of galaxies}},  {\em JCAP} {\bf 01} (2017) 032,
  [\href{http://arxiv.org/abs/1512.03918}{{\tt arXiv:1512.03918}}].

\bibitem{Beutler:2018vpe}
F.~Beutler, E.~Castorina, and P.~Zhang, {\it {Interpreting measurements of the
  anisotropic galaxy power spectrum}},  {\em JCAP} {\bf 03} (2019) 040,
  [\href{http://arxiv.org/abs/1810.05051}{{\tt arXiv:1810.05051}}].

\bibitem{Beutler:2020evf}
F.~Beutler and E.~Di~Dio, {\it {Modeling relativistic contributions to the halo
  power spectrum dipole}},  {\em JCAP} {\bf 07} (2020), no.~07 048,
  [\href{http://arxiv.org/abs/2004.08014}{{\tt arXiv:2004.08014}}].

\bibitem{McDonald:2009ud}
P.~McDonald, {\it {Gravitational redshift and other redshift-space distortions
  of the imaginary part of the power spectrum}},  {\em JCAP} {\bf 11} (2009)
  026, [\href{http://arxiv.org/abs/0907.5220}{{\tt arXiv:0907.5220}}].

\bibitem{Irsic:2015nla}
V.~Ir\v{s}i\v{c}, E.~Di~Dio, and M.~Viel, {\it {Relativistic effects in
  Lyman-\ensuremath{\alpha} forest}},  {\em JCAP} {\bf 02} (2016) 051,
  [\href{http://arxiv.org/abs/1510.03436}{{\tt arXiv:1510.03436}}].

\bibitem{DiDio:2018zmk}
E.~Di~Dio and U.~Seljak, {\it {The relativistic dipole and gravitational
  redshift on LSS}},  {\em JCAP} {\bf 04} (2019) 050,
  [\href{http://arxiv.org/abs/1811.03054}{{\tt arXiv:1811.03054}}].

\bibitem{Bonvin:2023jjq}
C.~Bonvin, F.~Lepori, S.~Schulz, I.~Tutusaus, J.~Adamek, and P.~Fosalba, {\it
  {A case study for measuring the relativistic dipole of a galaxy
  cross-correlation with the Dark Energy Spectroscopic Instrument}},  {\em Mon.
  Not. Roy. Astron. Soc.} {\bf 525} (2023), no.~3 4611--4627,
  [\href{http://arxiv.org/abs/2306.04213}{{\tt arXiv:2306.04213}}].

\bibitem{Planck:2018vyg}
{\bf Planck} Collaboration, N.~Aghanim et~al., {\it {Planck 2018 results. VI.
  Cosmological parameters}},  {\em Astron. Astrophys.} {\bf 641} (2020) A6,
  [\href{http://arxiv.org/abs/1807.06209}{{\tt arXiv:1807.06209}}]. [Erratum:
  Astron.Astrophys. 652, C4 (2021)].

\bibitem{Etherington}
I.~M.~H. {Etherington}, {\it {On the Definition of Distance in General
  Relativity.}},  {\em Philosophical Magazine} {\bf 15} (1933) 761.

\bibitem{Yoo:2016vne}
J.~Yoo and F.~Scaccabarozzi, {\it {Unified Treatment of the Luminosity Distance
  in Cosmology}},  {\em JCAP} {\bf 09} (2016) 046,
  [\href{http://arxiv.org/abs/1606.08453}{{\tt arXiv:1606.08453}}].

\bibitem{Nadolny:2021hti}
T.~Nadolny, R.~Durrer, M.~Kunz, and H.~Padmanabhan, {\it {A new way to test the
  Cosmological Principle: measuring our peculiar velocity and the large-scale
  anisotropy independently}},  {\em JCAP} {\bf 11} (2021) 009,
  [\href{http://arxiv.org/abs/2106.05284}{{\tt arXiv:2106.05284}}].

\bibitem{Maartens:2021dqy}
R.~Maartens, J.~Fonseca, S.~Camera, S.~Jolicoeur, J.-A. Viljoen, and
  C.~Clarkson, {\it {Magnification and evolution biases in large-scale
  structure surveys}},  {\em JCAP} {\bf 12} (2021), no.~12 009,
  [\href{http://arxiv.org/abs/2107.13401}{{\tt arXiv:2107.13401}}].

\bibitem{Dalang:2021ruy}
C.~Dalang and C.~Bonvin, {\it {On the kinematic cosmic dipole tension}},  {\em
  Mon. Not. Roy. Astron. Soc.} {\bf 512} (2022), no.~3 3895--3905,
  [\href{http://arxiv.org/abs/2111.03616}{{\tt arXiv:2111.03616}}].

\bibitem{Scaccabarozzi:2018vux}
F.~Scaccabarozzi, J.~Yoo, and S.~G. Biern, {\it {Galaxy Two-Point Correlation
  Function in General Relativity}},  {\em JCAP} {\bf 10} (2018) 024,
  [\href{http://arxiv.org/abs/1807.09796}{{\tt arXiv:1807.09796}}].

\bibitem{Grimm:2020ays}
N.~Grimm, F.~Scaccabarozzi, J.~Yoo, S.~G. Biern, and J.-O. Gong, {\it {Galaxy
  Power Spectrum in General Relativity}},  {\em JCAP} {\bf 11} (2020) 064,
  [\href{http://arxiv.org/abs/2005.06484}{{\tt arXiv:2005.06484}}].

\bibitem{Castorina:2021xzs}
E.~Castorina and E.~Di~Dio, {\it {The observed galaxy power spectrum in General
  Relativity}},  {\em JCAP} {\bf 01} (2022), no.~01 061,
  [\href{http://arxiv.org/abs/2106.08857}{{\tt arXiv:2106.08857}}].

\end{thebibliography}\endgroup
\bibliographystyle{JHEP}
\end{document}